\documentclass[aps,pre,twocolumn,showpacs,amsmath,,amssymb,amsfonts,notitlepage]{revtex4-1}
\usepackage{graphicx}
\usepackage{dcolumn}
\usepackage{bm}
\usepackage{color}
\usepackage{multirow}

\usepackage{hyperref}
\pdfoutput=1

\newcommand{\rom}[1]{\uppercase\expandafter{\romannumeral #1\relax}}

\begin{document}
\title{Front propagation versus bulk relaxation in the annealing dynamics of a kinetically constrained model of ultrastable glasses}

\author{Ricardo Guti\'errez and Juan P. Garrahan}
\affiliation{School of Physics and Astronomy, University of Nottingham, Nottingham, NG7 2RD, UK}

\date{\today}
\keywords{}
\begin{abstract}
Glasses prepared by physical vapour deposition have been shown to be remarkably more stable than those prepared by standard cooling protocols, with properties that appear to be similar to systems aged for extremely long times.  When subjected to a rapid rise in temperature, ultrastable glasses anneal towards the liquid in a qualitatively different manner than ordinary glasses, with the seeming competition of different time and length scales.  We numerically reproduce the phenomenology of ultrastable glass annealing with a kinetically constrained model, a three dimensional East model with soft constraints, in a setting where the bulk is in an ultrastable configuration and a free surface is permanently excited.  Annealing towards the liquid state is given by the competition between the ballistic  propagation of a front from the free surface and a much slower nucleation-like relaxation in the bulk.  The crossover between these mechanisms also explains the change in behaviour with film thickness seen experimentally. 
\end{abstract}

\maketitle

\section{Introduction}

Glasses are out-of-equilibrium systems, and as such their properties are history dependent \cite{Struik1977,Ediger1996,Angell2000,Binder2011}. The standard preparation method for a large class of glass formers consists in cooling down an equilibrated liquid below the melting point in such a way that the metastable supercooled liquid does not crystallise, decreasing its temperature further until the relaxation time exceeds the experimental time scales at the (protocol-dependent) glass transition temperature $T_g$. In recent years, however, an alternative preparation technique \cite{swallen2007,kearns2009,swallen2010,dalal2012,sepulveda2013} has been shown to have interesting applications in materials science and has provided the physics of the glass transition with a new set of intriguing experimental facts whose theoretical elucidation may help increase our understanding of the glass transition more generally. 

This technique is physical vapour deposition, which consists in the addition of layers of molecules onto a substrate at a low temperature $T_\textrm{dep}$. Vapour deposition produces systems characterised by unusual thermodynamic and kinetic stability properties when $T_\textrm{dep}$ is moderately smaller than $T_g$, especially around the apparently optimal temperature $T_\textrm{dep} \approx 0.85\, T_g$, which have earned these systems the name of {\it ultrastable glasses} \cite{swallen2007}. Stable glasses show, when compared to ordinary glasses resulting from cooling a liquid, lower enthalpies and higher onset temperatures \cite{swallen2007}, higher densities and lower fictive temperatures \cite{dalal2012}. According to these studies, the kinetic and thermodynamic properties of stable glasses are similar to those one would expect from ordinary glasses after a very prolonged period of aging. These systems are thus believed to lie in very deep regions of the potential energy landscape that can only be reached by ordinary means after a good deal of rearrangement of local configurations (which is of necessity extremely sluggish at such low temperatures).

The stability of ultrastable glasses can be tested by studying how they revert back to the liquid state when annealed above the experimental glass transition temperature. A key experimental observation is that stable glasses do not transform to the liquid in the same manner as ordinary glasses: at least in thin enough vapour-deposited films, melting takes place first in the vicinity of the free surface, and then propagates at constant speed to the rest of the system \cite{swallen2009,swallen2010,rodriguez2014}. This is further confirmed by the fact that the inclusion of additional planes of mobility generates new propagation fronts starting from each of them \cite{sepulveda2013}. The propagation speed of the front and its dependence on the structural relaxation time at the annealing temperature and on the deposition temperature has been characterised in a number of recent publications \cite{rodriguez2014,rodriguez2015,tylinski2015}. Furthermore, a recently experimental study has analysed the enhanced dynamics of ultra-thin vapour-deposited glass films, revealing a strong correlation between the dynamics of the free surface and the bulk over considerably long scales \cite{zhang2016}.
A second key observation is that the time scale associated with transforming an ultrastable film into the liquid displays a crossover with film thickness, from a linear dependence on thickness for thin films to becoming independent of thickness at some threshold size \cite{kearns2009,sepulveda2014}.  This suggests a competition between relaxation dynamics initiated at the free surface (and, if they exist, other planes of higher mobility - including possibly the glass-substrate interface \cite{swallen2010,sepulveda2013}) and transformation processes initiated in the bulk of the system \cite{swallen2009,kearns2009,sepulveda2014}.

In this paper we consider the problem of melting of ultrastable glasses from the dynamical facilitation point of view \cite{Chandler2010}.  Specifically, we study numerically the annealing of ultrastable glass films modelled by means of a three-dimensional East facilitated spin model \cite{Ritort2003,Ashton2005,Berthier2005,Chleboun2014,Chleboun2014b,Chleboun2015} (also known as North-or-East-or-Front model) with soft constraints \cite{Elmatad2013}, a model known to display many features associated with glassy dynamics such as super-Arrhenius relaxation and dynamic heterogeneity.  Several properties of ultrastable glasses have already been considered using kinetically constrained \cite{Ritort2003}
or associated plaquette models \cite{Garrahan2002}
(for studies of stable glasses with other approaches see e.g.\ \cite{Wolynes2009,Singh2011,Wisitsorasak2013,Mirigian2014}).  This includes ultrastable glass preparation and aspects of their relaxation when modelled by a three-spin-facilitated 
Fredrickson-Andersen model \cite{leonard2010,douglass2013}, or nucleation and growth dynamics in their bulk relaxation and the relation to overlap transitions in coupled plaquette models  \cite{Jack2016} .  Here we specifically study the competition between surface and bulk relaxation mechanisms, which to our knowledge has only been observed experimentally to date \cite{kearns2009,sepulveda2014}. 
This is a crucial aspect of the transformation dynamics of stable glasses, as the difference in the way stable glasses and ordinary glasses transform into the liquid can be attributed to the presence of these mechanisms and to a crossover between them.  As we show below, our model qualitatively replicates the phenomenology observed experimentally \cite{sepulveda2014}, indicating that the dynamics of stable glasses can be rationalised with dynamic facilitation ideas.

\section{Model}

In order to model ultrastable glasses in the simplest possible way we will consider a three-dimensional version of the East model (or North-or-East-or-Front model) \cite{Ritort2003,Ashton2005,Berthier2005,Chleboun2014,Chleboun2014b,Chleboun2015}.  The East model and its generalisations are known to display many of the dynamical characteristics both of supercooled liquids, including super-Arrhenius relaxation times in equilibrium \cite{Sollich1999,Chleboun2014} (with a ``parabolic'' law that works well phenomenologically \cite{Elmatad2009}), dynamical heterogeneity \cite{Garrahan2002a}, transport decoupling \cite{Jung2004,Blondel2014}, and of glasses, such as anomalous thermodynamic responses when driven out of equilibrium \cite{Keys2013}.  Furthermore, the East model and its generalisations, while highly non-trivial, are simple enough to allow for systematic studies, and many properties of their dynamics are known rigorously \cite{Faggionato2012,Blondel2013,Chleboun2014,Chleboun2014b,Chleboun2014c}.

The model of a liquid or glass film that we study consists of a system of Ising spins on a cubic lattice of size $N = L \times L \times h$, where $L$ corresponds to the plane dimensions and $h$ to the vertical size of the film.  Like other facilitated spin models, the energy function of the system is non-interacting, $E = \sum_i^N n_i$, where $n_i = 0,1$ indicates the state of spin at site $i$. 
The dynamics is kinetically constrained in a way that we now explain.  In the 3D East model the (hard) kinetic constraint on spin $i$ at position $(x_i,y_i,z_i)$ is given by the binary variable $c_i$, which is dependent on the local configuration: $c_i = 1$ (i.e., spin $i$ can flip) if at least one of the spins at $(x_i+1,y_i,z_i)$, $(x_i,y_i+1,z_i)$ and $(x_i,y_i,z_i+1)$ is excited, and $c_i=0$ (i.e., spin $i$ cannot flip) otherwise. [Periodic boundary conditions are considered along the $x$ and $y$ directions, but not along the $z$ direction, so that $(x_i,y_i,z_i+1)$ is only inspected if $z_i < h$.]  For reasons that are explained in detail below, in our modelling we will consider a ``soft'' constraint.  For a given inverse temperature $\beta = 1/T$ (in units such that $k_\textrm{B} = 1$), an excitation process at a generic site $i$, $n_i = 0 \to 1$, occurs with an associated rate $\left[c_i + \exp{(-\beta U)}\right]\exp{(-\beta)}$,  and a de-excitation process, $n_i = 1 \to 0$, occurs with a rate $c_i + \exp{(-\beta U)}$. The configuration-independent rate $\exp{(-\beta U)}$ thus softens the constraint by allowing for the spontaneous occurrence of spin flips in the absence of neighbouring excitations with an associated energy cost $U$ \cite{Elmatad2013}. This dynamics satisfies detailed balance with respect to the canonical distribution $\exp(-\beta E[\{n_i\}])/\mathcal{Z}$ for the non-interacting Hamiltonian $E$ above.  Due to the kinetic constraints, the dynamics is much richer than the underlying thermodynamics \cite{Ritort2003}, which is the central aspect of the dynamic facilitation approach \cite{Chandler2010}. 

\begin{figure}[t]
\begin{flushleft}
\includegraphics[scale=0.267,natwidth=4000,natheight=1795]{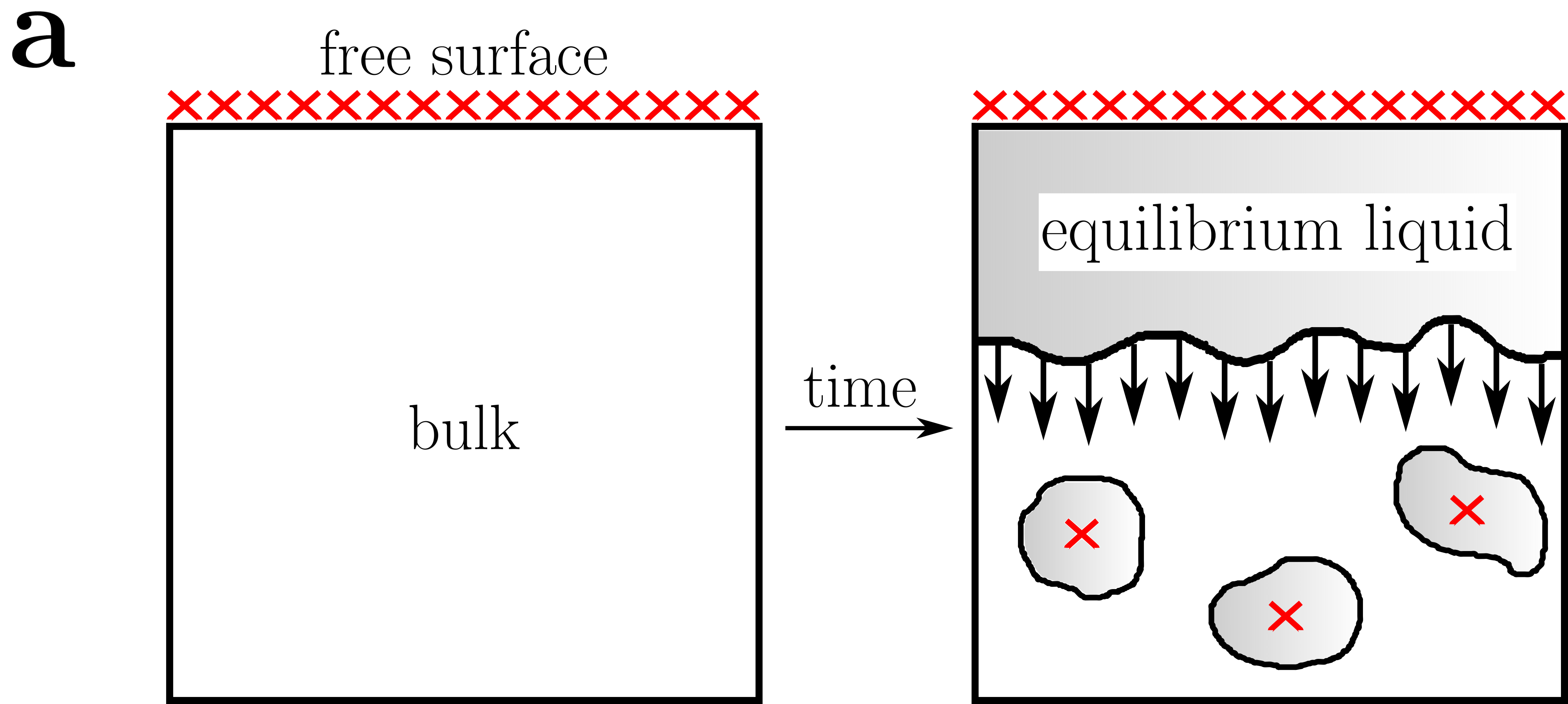}\\
\vspace{0.7cm}
\includegraphics[scale=0.109,natwidth=2271,natheight=1271]{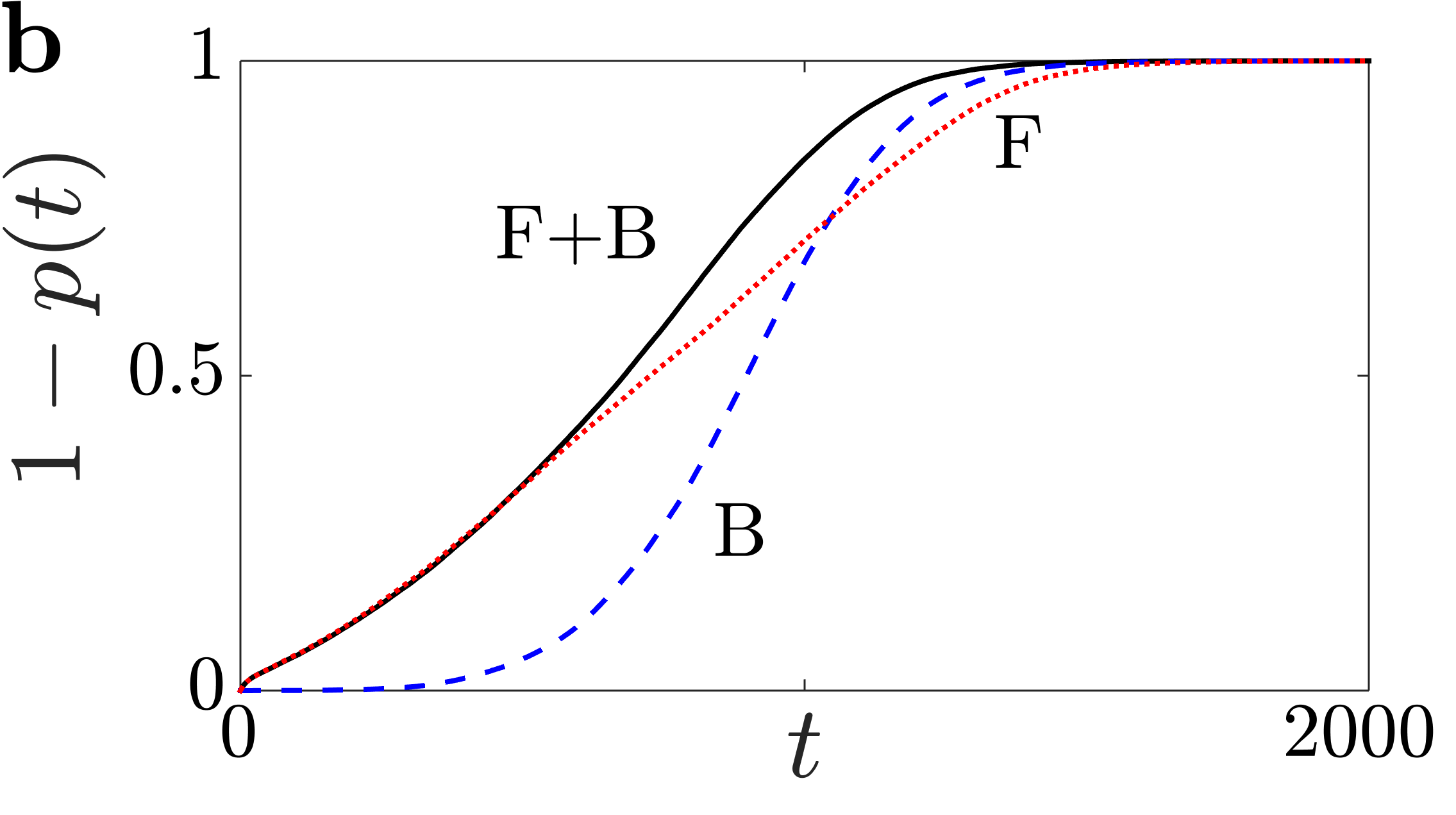}
\end{flushleft}
\vspace{-0.5cm}
\caption{ {\sf \bf Illustration of the transformation dynamics, including a front propagating from the free surface and a nucleation and growth process in the bulk.} (a) Sketch of the system. Initially, there is a top layer of excitations, which are fixed throughout time to reflect the enhanced mobility of the free surface in experimental systems, and an empty bulk. As time proceeds, an excitation front propagates from the top layer and new isolated excitations are sponatenously created (nucleate) and grow in the bulk; cf.\ Ref.\ \cite{kearns2009}.  Red crosses indicate the free surface excitations and the spontaneously created excitations in the bulk, while the grey areas adjacent to them correspond to the regions that become excited through local facilitation.     (b)  Transformation dynamics of a $64 \times 64 \times 64$ lattice at $T=0.45$ as given by the fraction of spins that have flipped since the initial time $1 - p(t)$ in three different situations: (F) front propagation, (B)  bulk nucleation and growth, (F+B)  both mechanisms combined. (The parameter choices used in each case are given in the text.) Each curve results from averaging $20$ independent realisations.}
\label{fig1}
\end{figure}

Our setup for the annealing dynamics is sketched in Fig.\ \ref{fig1}(a): The top layer ($z=h$) of the simulation box is permanently excited, as it represents the highly mobile free surface of the vapour-deposited system, while the bulk is initially without excitations, cf.\ Ref.\ \cite{leonard2010}.  
The initial configuration sketched on the left panel corresponds to the bulk of
an ultrastable glass of low fictive temperature.  As such the bulk concentration of excitations is much smaller that the one at equilibrium at the annealing temperature $T$, $c_\textrm{eq} = (1 + \exp{\beta})^{-1}$. As time proceeds, the free surface gives rise to neighbouring excitations in the layers immediately underneath.  This results in a relaxation front that proceeds inwards from the free surface.  We expect the front to advance ballistically on average, leaving behind it an equilibrium concentration of excitations (i.e.\ the equilibrated ``liquid'').  This ballistic propagation with equilibrium behind was indeed proven for the one-dimensional East model \cite{Blondel2013}.  A second relaxation mechanism originates in the bulk, where facilitated dynamics is initially not possible, while the spontaneous creation of excitations is suppressed but not completely absent.  An excitation can indeed be spontaneously created 
with a rate $\exp{(-\beta U)}$ (due to the softness of the constraint), which can then facilitate further excitations in their vicinity.  This is therefore a nucleation and growth process.  The relative time scales over which theses two processes - front propagation from the free surface and bulk excitation - occur may give rise to different annealing dynamics.

Indeed, the language of nucleation theory is relevant in this context, and we will see below that the Kolmorogov-Johnson-Mehl-Avrami theory of nucleation and growth \cite{kolmogorov1937,johnson1939,avrami1939,avrami1940,avrami1941} applies to bulk relaxation in our setup.  This connection already appeared in some experimental papers \cite{kearns2009}, and has been used recently in the theoretical study of bulk relaxation of ultrastable glasses with plaquette models \cite{Jack2016}.  In this sense, the softness of the constraint can be thought of as a simplified version of the nucleation mechanism described in Ref.\ \cite{Jack2016} in the context of plaquette models (which as effective facilitated models also have soft constraints).  We will elaborate on this point later in the paper.  The time scales over which front propagation and spontaneous excitation occur will depend on both the annealing temperature $T$ and the soft constraint barrier height $U$. The latter is the effective parameter in our model that controls the stability of the stable glass.  In experiments this is related to the deposition temperature of the stable glass $T_\textrm{dep}$ (i.e. the closer $T_\textrm{dep}$ is to the optimal deposition temperature, the larger the effective barrier height $U$).

\section{Annealing dynamics: front propagation vs. bulk relaxation}

With our simplified model, we aim to elucidate the annealing process whereby an ultrastable glass turns into a liquid. This transformation reveals potentially interesting properties of stable glasses, and the structural aspects in which they differ from ordinary glasses, and has been carefully studied in experiments \cite{kearns2009, sepulveda2014}. The transformation into the liquid state in those studies is inferred from changes in the specific heat  \cite{kearns2009} or in the dielectric loss response  \cite{sepulveda2014}. 

In our case, a simple observable that can give an estimate of the amount of material that has transformed into liquid is given by the fraction of spins that have 
flipped at least once from their initial state, i.e.  $1-p(t)$, where $p(t)$ is the persistence function, see e.g.\ \cite{Ashton2005}.  By exploring the behaviour of $1-p(t)$ for different values of $T$ (annealing temperature), $U$ (which acts as a proxy for the fictive temperature or the glass), and $h$ (film thickness) we expect to shed light on the mechanisms underlying the melting of stable glasses.

An illustration of the competition at the heart of the annealing dynamics is given in Fig.\ \ref{fig1}(b).  It shows the relaxation of a system of $N = 64 \times 64 \times 64$ sites at $T=0.45$,  in three different situations: the curve denoted (F) corresponds the case where the melting dynamics is solely due to the propagation of the front from the free surface ($U \to \infty$); (B) corresponds to pure bulk relaxation ($U=6$, no free surface, and periodic boundary conditions in the $z$ direction); and (F+B) to a situation when both mechanisms are at play, so that there is an initial front propagation before bulk relaxation takes over ($U=6$ with a free surface).
In this way (F) and (F+B) represent stable glasses of extreme or finite stability, respectively, while (B) shows the relaxation of the bulk in the absence of a free surface of excitations \cite{douglass2013}.  We see from Fig.\ \ref{fig1}(b) that the shape of the (F+B) curve can be explained by a combination of the (F) and (B) processes [it cannot be just the sum, as a spin that has already been flipped by bulk relaxation or front propagation is already liquid-like, and will not increase its contribution to $1-p(t)$ at the time the other process makes it flip]. With this illustration in mind, we now turn our attention to a study of the front dynamics, then of the bulk dynamics, and finally, in the next section, to the combination of both with special attention to the time scales involved.

\begin{figure}[t]
\includegraphics[scale=0.11,natwidth=2000,natheight=1262]{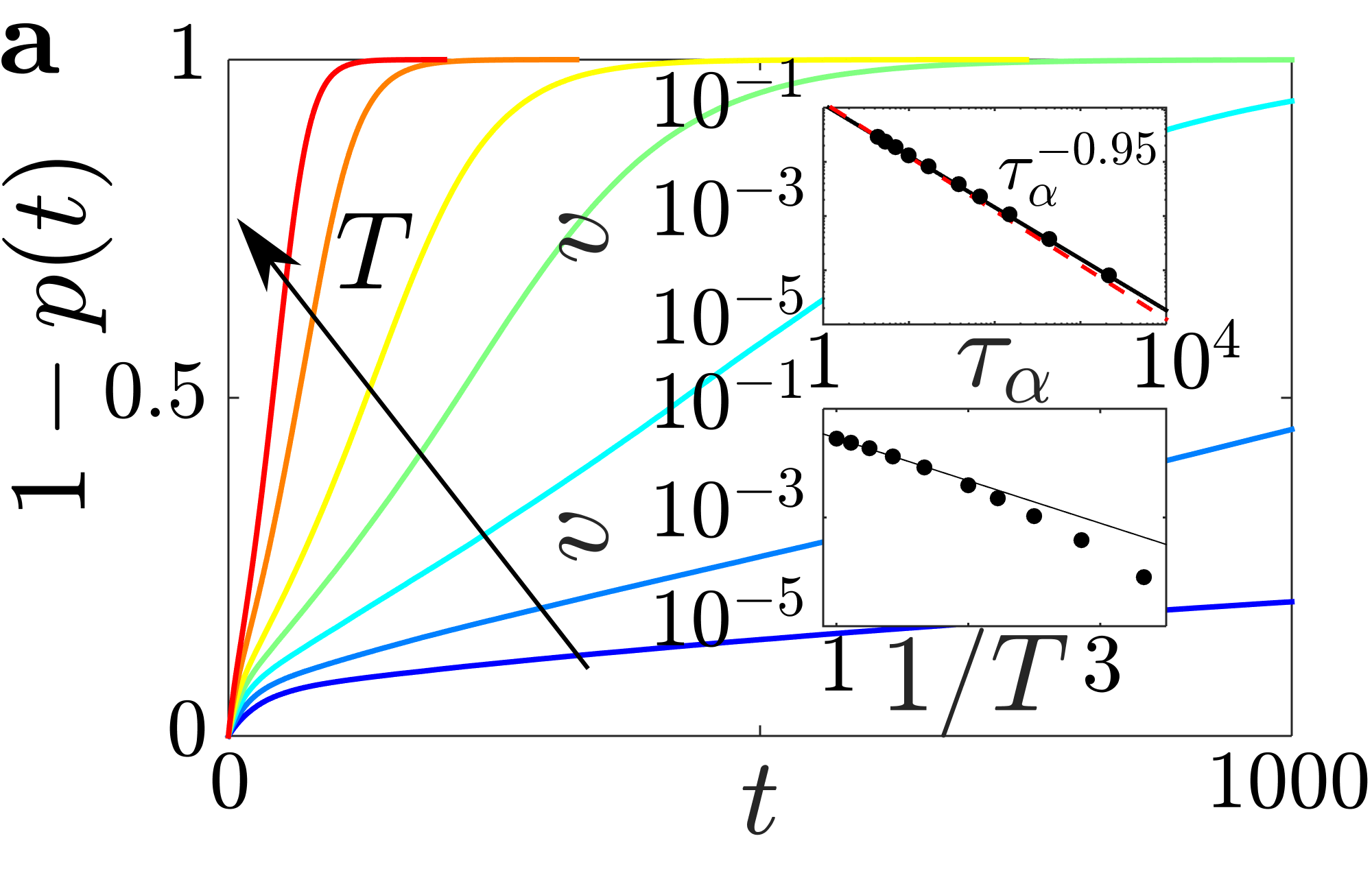}
\includegraphics[scale=0.11,natwidth=2000,natheight=1262]{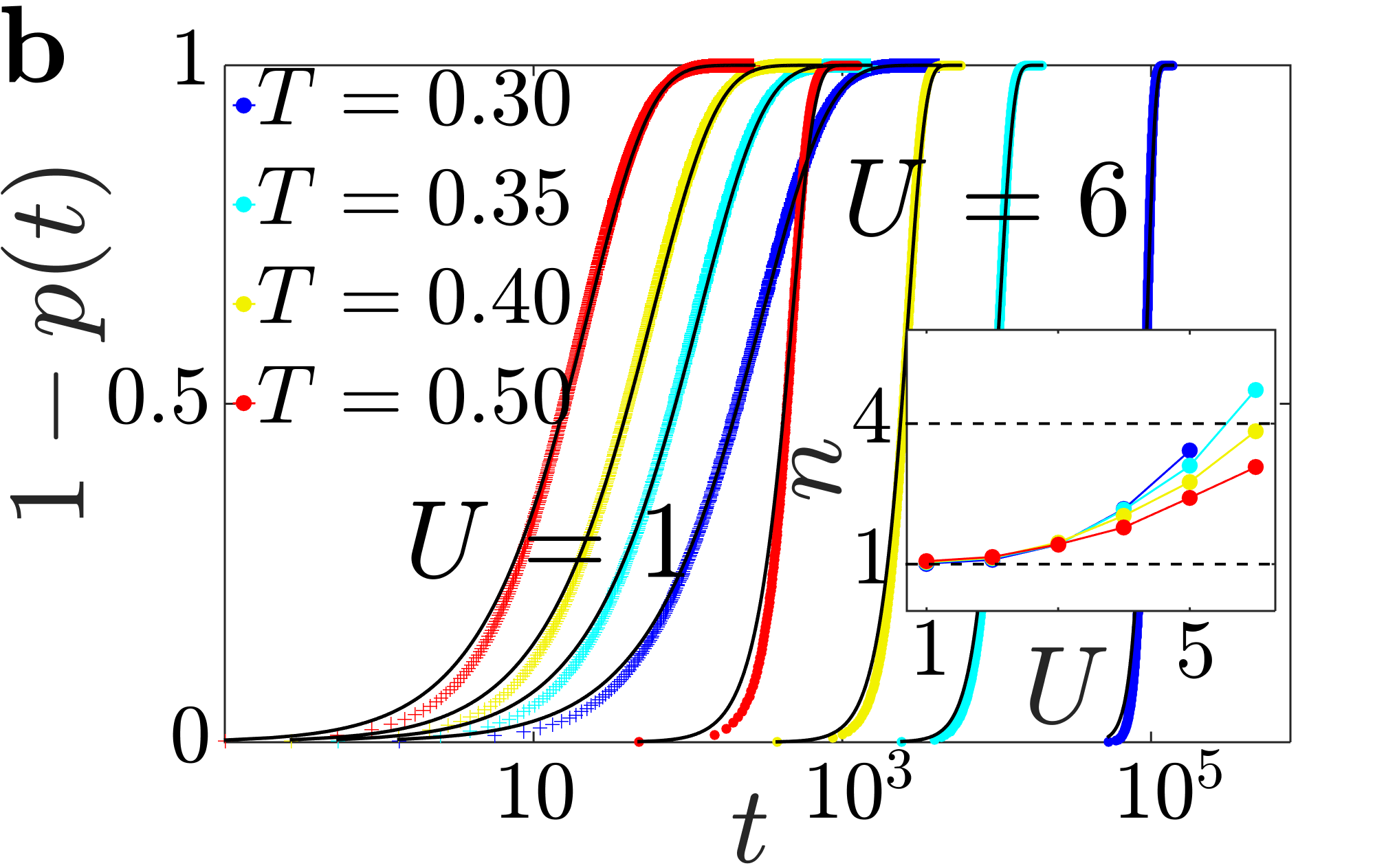}
\caption{ {\sf \bf Front propagation and bulk relaxation.} (a)  Annealing dynamics of a system of size $64\times 64\times 16$ for $T=0.30, 0.35, 0.40, 0.45, 0.50, 0.60, 0.70$ and $U\to\infty$. Blue corresponds to low temperatures and red to high temperatures, and the temperature grows in the direction of the black arrow. Lower inset: front propagation speed $v$ as a function of $1/T$. The line is just a visual aid to highlight the super-Arrhenius behaviour. Upper inset: $v$ as a function of $\tau_\alpha$, where $\tau_\alpha$ is the structural relaxation time of an equilibrated system in the absence of a free surface. (b) Transformation dynamics of a $64\times 64\times 64$ system without free surface for $T=0.30,0.35,0.40$ and $0.50$ (see colour coding in the legend), for $U=1$ (left, crosses) and $U=6$ (right, circles). Fits based on the Avrami equation $1-p(t) = 1 - \exp(-k\, t^n)$ are also included (see black lines). Inset: Avrami exponents $n$ resulting from these fits as functions of $U$ for different $T$. In all panels, each curve is an average of $20$ independent realisations.}
\label{fig2}
\end{figure}

The propagation of the front originating from the free surface seems to occur at a constant speed throughout the annealing process, except for the very initial and very late stages.  Ballistic spreading from the free surface has been proven to occur in the one-dimensional East model \cite{Blondel2013}, and we expect it also to hold in higher dimensions (and was seen to occur as well in the three-spin facilitated Fredrickson-Andersen model in \cite{leonard2010}).  A key aspect is that as the front propagates it leaves behind equilibrated configurations \cite{Blondel2013,Chleboun2014}, so that the front passing through is enough to transform the glass into the liquid.  Our detailed results are given in Fig.~\ref{fig2}(a), where we show the annealing dynamics of a system of $N=64\times 64\times 16$ for $T=0.30, 0.35, 0.40, 0.45, 0.50, 0.60, 0.70$ and $U\to\infty$. Blue lines correspond to low temperatures and red lines to high temperatures, with the temperature growing in the direction of the black arrow. The fact that we have taken a system of only $h=16$ spins in the $z$ direction instead of $h=64$ for computational reasons makes it more evident than in Fig.~\ref{fig1}(b) that there is an initial stage of nonlinear behaviour, but later on the annealing dynamics follows extremely closely a straight line until the system is almost fully transformed. In the lower inset of Fig.~\ref{fig2}(a) we show the front propagation speed $v$ as a function of $1/T$. The fact that $v(1/T)$ deviates from a straight line in semilog scale for larger $1/T$ means that the propagation time $v^{-1}$ grows faster than $\exp(\Delta E/T)$, for a certain barrier $\Delta E$, as the temperature is decreased, indicating super-Arrhenius behaviour in the front propagation.This is in agreement with recent experimental results \cite{rodriguez2015}. 

In the upper inset of Fig.~\ref{fig2}(a), we plot $v$ as a function of the structural relaxation time $\tau_\alpha$ for different $T$ in loglog scale. The relaxation time $\tau_\alpha$ is extracted from the decay of the persistence to $e^{-1}$ in an equilibrated system in the absence of a free surface (with periodic boundary conditions).  As we are considering $U\to\infty$, this corresponds to the relaxation time of the North-or-East-or-Front model with hard constraints, which is known to grow as the temperature is decreased in a super-Arrhenius fashion as well \cite{Ashton2005,Berthier2005}. If the structural relaxation controlled the front propagation we would expect $v \sim \tau_\alpha^{-\gamma}$ for $\gamma=1$ \cite{tylinski2015}. However, we find an exponent of $\gamma = 0.95$, which lies close but is still visibly distinct from the $\gamma=1$ case, to give the best fit to our data (see the black continuous line; the red dashed line corresponds to $\gamma = 1$ and has been included for comparison). Indeed, the relation between $v$ and $\tau_\alpha$ has been explored experimentally using different glass formers, and in all cases the results seem to be compatible with a power law relation with $\gamma < 1$  \cite{rodriguez2014,tylinski2015}. Recently, the dependence of this relation on the deposition temperature has been investigated, with results that indicate that the exponent $\gamma$ is independent of the preparation, while the prefactor becomes smaller as the stability of the sample increases \cite{rodriguez2015}. This form of decoupling has not only been observed in experimental systems, but also in the kinetically constrained model of  Ref.~\cite{leonard2010}, where a front propagation phenomenology qualitatively similar to the one we report in this work was first studied. While those works show an exponent $\gamma$ that ranges between $0.7$ and $0.9$, our value seems compatible with a milder form of decoupling. Further research is required to elucidate whether this is simply due to the relatively weak kinetic constraint or other peculiarities of the three-dimensional East model, and its possible relation to the essentially compatible with $\gamma=1$ results found in the melting of ultrastable glasses arising from random pinning \cite{hocky2014}.

\begin{figure*}[t]
\begin{minipage}[b]{0.3295\textwidth}
\includegraphics[scale=0.085,natwidth=2000,natheight=1262]{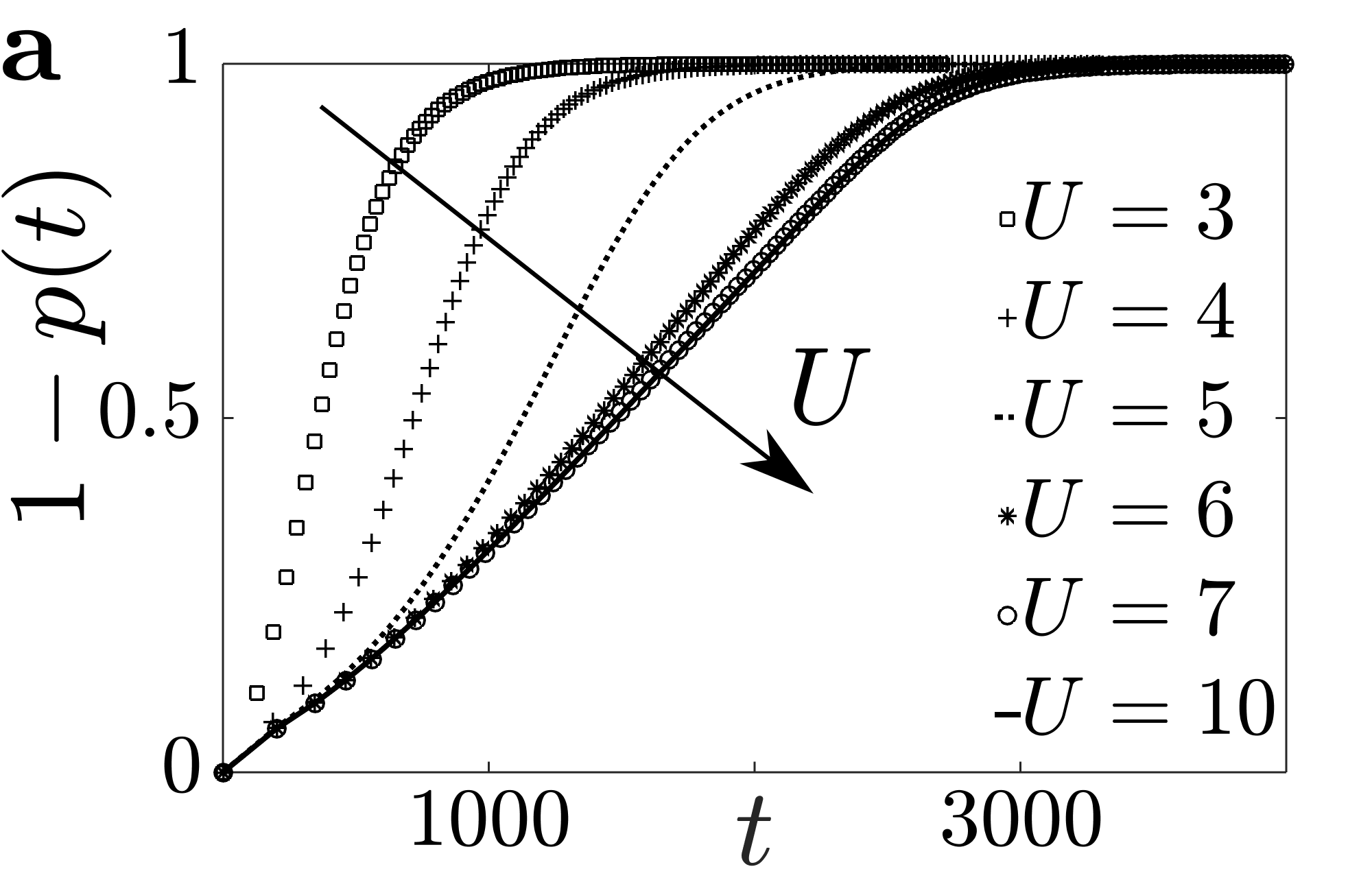}
\end{minipage}
\begin{minipage}[b]{0.3295\textwidth}
\includegraphics[scale=0.085,natwidth=2000,natheight=1262]{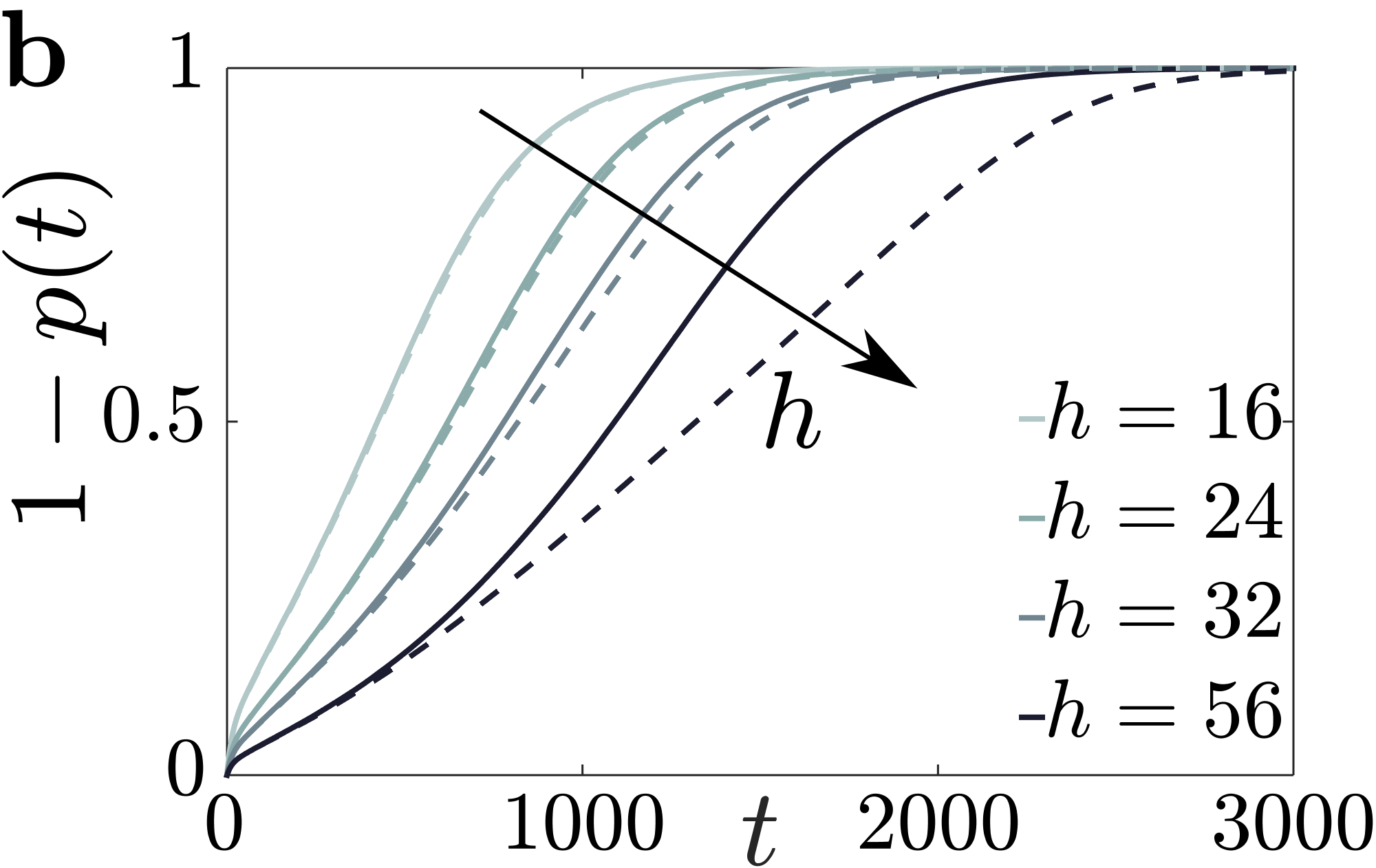}
\end{minipage}
\begin{minipage}[b]{0.3295\textwidth}
\includegraphics[scale=0.0855,natwidth=2000,natheight=1271]{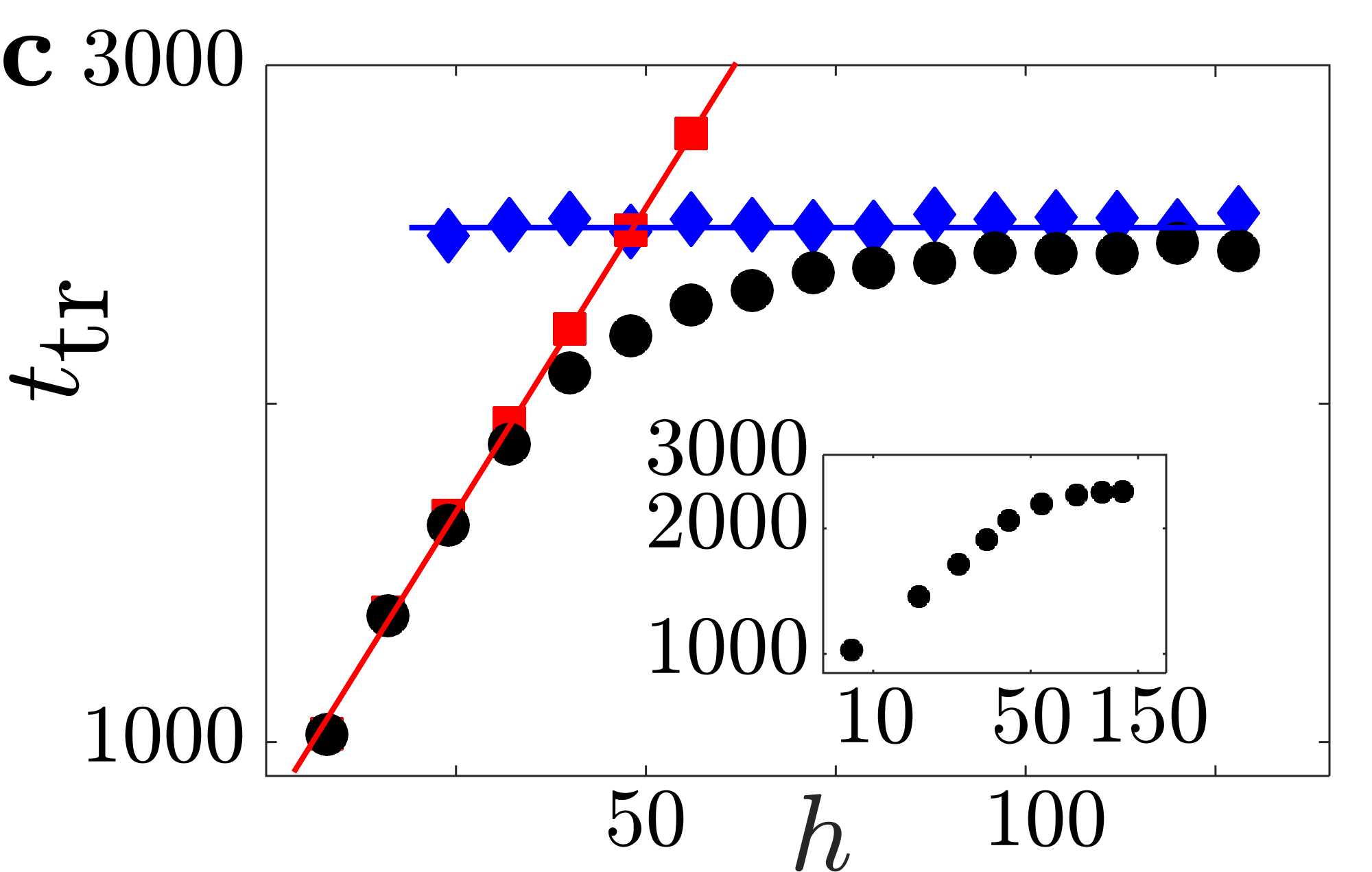}
\end{minipage}
\caption{ {\sf \bf Annealing dynamics and transformation times for different values of $T$ and $U$.} (a) Transformation into the liquid of a $64 \times 64 \times 64$ lattice at $T=0.4$ for different values of $U$. The fraction of spins that have flipped since the initial time $1 - p(t)$ is shown for $U = 3,4,5,6,7,10$. (b) Transformation into the liquid for $T=0.4$ and $U=5$, in a system of $64\times 64\times h$ spins, for different values of $h$. The continuous lines correspond to $U=5$, whereas the dashed lines correspond to $U\to\infty$ (pure front propagation). (c) Transformation time $t_\textrm{tr}$ for different $h$ with parameters as in (b). We define $t_\textrm{tr}$ as the time at which $99\%$ of the sample has transformed into the liquid. Results for $U=5$ with (black circles) and without a free surface (blue diamonds), and $U\to\infty$ with a free surface (red squares) are included in the plot. Colour coding as in Fig.~\ref{fig1} (b). All results are based on averages of $20$ independent realisations.}
\label{fig3}
\end{figure*}

We next study the bulk dynamics. In this case, there is no free surface of excitations, and we are not only interested in the dependence of our results on $T$, but also on $U$, which is now finite. In Fig.~\ref{fig2}(b) we show the transformation dynamics of a $N=64\times 64\times 64$ system without free surface for $T=0.30,0.35,0.40$ and $0.50$ (see colour coding in the legend), for $U=1$ (left, crosses) and $U=6$ (right, circles). As in the experimental data reported in \cite{kearns2009}, the bulk evolution can be fit to an Avrami form \cite{avrami1939}, $1-p(t) = 1 - \exp(-k\, t^n)$, which as shown in the figure (see the black lines) fits the data fairly well. The Avrami exponents $n$ resulting from these fits are shown in the inset. For small $U$ the growth of the nuclei is irrelevant in the sense that it is slower than the nucleation of new excitations, and this is reflected in the fact that $n\approx 1$ \cite{christian2002}. For larger values of $U$, the exponent grows and gets closer to the value $n=4$ proposed in the Kolmorogov-Johnson-Mehl-Avrami theory of phase transformations when the growth is isotropic and linear in time \cite{kolmogorov1937,johnson1939,avrami1939,avrami1940,avrami1941,christian2002}. While for values such as $U=1, 2$ or $3$, the Avrami equation closely matches the experimental data, for larger values of $U$ the fit becomes worse (as seen when the results are plotted in linear time scale, or by the mean squared error of the least-squares fit procedure, not shown here). Indeed, the exponent $n>4$ that we obtain for $T=0.35$ and $U=6$ probably originates from an insufficiently good fit of the Avrami equation to the data, and the exponent for $T=0.30$ and $U=6$ is not even included in the inset as the fit looks particularly bad in that case. The origin of this discrepancy, which becomes conspicuous for small $T$ and large $U$ values (i.e. for systems with very small nucleation rates), is most probably a finite-size effect, arising when the typical distance between newly created excitations is of the order of the linear size of the system. This hypothesis is vindicated by the fact that performing the same analysis on substantially smaller systems for a given temperature leads to poorer Avrami fits for small values of $U$ that are not problematic at all in these relatively larger systems.

\section{Annealing dynamics: connection to experimental results}

Having considered separately the front and bulk dynamics and their dependence on $T$ and $U$, we next turn to the dynamics of the full system, i.e.\ 
one with both a free surface and a finite $U$, where both mechanisms are at play, cf.\ (F+B) in Fig.~\ref{fig1}(b).  
The combination of the two relaxation routes in the model does indeed reflect the melting phenomenology of experimental stable glasses, as both a constant speed propagation \cite{kearns2009,leonard2010,sepulveda2014}, and a bulk relaxation that follows closely an Avrami form \cite{kearns2009,Jack2016} have been observed in those systems.  In order to to replicate the full phenomenology reported in \cite{kearns2009,sepulveda2014} we need to consider the two mechanisms taking into account the relevant time scales.

Before we establish a comparison between our numerical results and previous experimental findings, it is important to reiterate that the annealing process in vapour-deposited stable glasses is strongly dependent on the deposition temperature $T_\textrm{dep}$, as this temperature controls the stability of the glass \cite{swallen2007}. In our model the role of  $T_\textrm{dep}$ is played by $U$, as a more densely packed highly stable configuration is expected to need to overcome a higher barrier to activate a nucleus in the bulk and give rise to an excitation capable of growth by the constrained dynamics. Highly stable glasses produced at the optimal  $T_\textrm{dep} \approx 0.85$ with very slow deposition rates would correspond to a very high $U$ in our model, whereas less stable films would correspond to smaller $U$ values. 

In Fig.~\ref{fig3}(a) we show the transformation dynamics for different values of $U$, in a system of $N=64\times 64 \times 64$ and $T=0.4$.  These results are analogous to those presented in Fig.\ 2 of Ref.\ \cite{sepulveda2014}, which is a detailed experimental study on the annealing dynamics of stable glasses. While the annealing is slowed down when the excitation barrier $U$ is increased, for sufficiently high barriers, $U\geq 7$, it becomes independent of the precise value of $U$. As the probability of creating a bulk excitation in a given finite time window in a finite-size system must vanish for sufficiently high $U$, this can only mean that above $U\approx 7$ (for this annealing temperature) the transformation is governed by the ballistic propagation of the excitation front originating in the free surface.  For smaller values of $U$ the dynamics has a contribution due to the nucleation and growth of excitations in the bulk, which is small for $U=6$, but large enough to obliterate any trace of a linear transformation for the smaller values of $U$ considered. The annealing dynamics in such cases becomes homogeneous, as in ordinary glasses.

An intriguing experimental observation  \cite{sepulveda2014} relates to the dependence of the total annealing time on the film height $h$, in particular the crossover from a linear dependence on $h$ for thinner films, to $h$-independent values for thicker films; see e.g.\ Fig.\ 3 of Ref. \cite{sepulveda2014}.  Furthermore, this crossover occurs at film heights of the order of a $\mu$m, an almost macroscopic length way beyond any expected dynamical correlation length scales of supercooled liquids at conditions near $T_{g}$.  By considering the height dependence in our model we argue that this micron sized length is not indicative of correlations. It is the size at which the two relaxation mechanisms crossover. 

Figure \ref{fig3}(b) (continuous lines) shows the ``liquid fraction'' as a function of time for various $h$. We use the same parameters as before, except that now $U=5$ is fixed and $h$ varies. To explicitly distinguish front from bulk relaxation we have included results for $U\to\infty$ (dashed lines), which show a relaxation purely based on a ballistic front propagation. When the system is thin ($h<32$), there is not enough time for the nucleation of excitations in the bulk, and the front dominates. For thicker samples, however, the front takes too long to reach the whole system, leaving enough time for the nucleation of bulk excitations to occur. At fixed annealing temperature $T$, the competition between both mechanisms is controlled by $U$ (in experiments, $T_\textrm{dep}$ and the deposition rate, as well as the microscopic properties of the glass former) and $h$: for highly stable glasses with large activation barriers, the bulk dominated transformation will only be seen in very thick samples, as it happens in the experimental results reported in Ref. \cite{sepulveda2014}.  

The competition between front propagation and bulk relaxation becomes evident in the manner in which the overall transformation time changes with height $h$. Here, the transformation time $t_\textrm{tr}$ is defined as the time at which the liquid fraction $1-p(t)$ reaches $0.99$.  This is shown in Fig.~\ref{fig3}(c) (black circles) for the same parameter values used in panel (b).  We
can see that for small $h$ the transformation time is linear in $h$, crossing over to a constant value at large $h$.  
For comparison we also show (blue diamonds) the transformation time from a pure bulk system (no free surface and p.b.c.\ in all directions, including $z$), and that due solely to a free surface (red squares, $U\to\infty$); cf.\ Fig.~\ref{fig1} (b), which contains the data used in the computation of the $h=64$ point in this panel.  For the conditions of Fig.~\ref{fig3}(c) the crossover occurs at $h_{\rm cross} \approx 50$, and in general this crossover length depends on $U$ (the stability of the glass) and $T$ (the annealing temperature). In the inset to Fig.~\ref{fig3}(c) we show the transformation times in loglog scale, which can be compared to the analogous experimental figures in Refs.\ \cite{kearns2009,sepulveda2014}.

\section{Conclusions}

This work sheds light on the melting of stable glasses using dynamic facilitation ideas. To this end, we have modelled ultrastable films using a three-dimensional East model with a free surface which is permanently excited, cf.\ \cite{leonard2010}.  We have shown that the competition between the relaxation initiated at the free surface, which propagates as a ballistic front that leaves behind equilibrated material, and bulk relaxation is responsible for the crossover behaviour seen in experiments.  A consequence of this competition is the emergence of a characteristic film thickness at which a front dominated melting regime crosses over to a bulk dominated melting regime, see Fig.\ \ref{fig3}(c).  This is the central result of this work.

An interesting aspect of the melting dynamics of stable glasses that we have not addressed is the manner in which this crossover length $l_c$ depends on the annealing temperature $T$ and on the stability of the sample. In practice the deviation from the linear growth illustrated in the (F+B) curve in Fig.~\ref{fig1} (b) only becomes visible when the fraction of glass that thas melted by bulk mechanisms is not negligibly small compared to the amount that has already equilibrated by the ballistic propagation of the front. The need for an unambiguous definition, however, requires the crossover length to be the length traversed by the front by the time the spontaneous creation of bulk excitations begins. As the average nucleation time is given by $\exp(U/T)$, the crossover length is therefore such that $l_c \sim v \exp(U/T)$. While in our work $v$ depends only on $T$, in experimental systems the front propagation is seen to become slower for more stable systems \cite{rodriguez2015}. As a larger stability also implies larger $U$, in principle $l_c$ may depend in a non-trivial way on the stability. A similar reasoning leads to the same conclusion in regard to the dependence on $T$. It seems that new experimental results that can be incorporated in the modelling of stable glasses may be needed to make progress in this direction.

In order to account for bulk relaxation having a super-exponential, or Avrami, time dependence we had to make the kinetic constraint soft. In this way the bulk of an ultrastable glass is modelled as devoid of excitations.  The absence of excitations prevents facilitated dynamics, but the softness of the constraint allows for rare spontaneous creation of excitations.  These assumptions are compatible with dynamic facilitation ideas about glasses: it is expected that in actual supercooled liquids effective kinetic constraints would be soft, with a small but non-vanishing probability of them being violated \cite{Garrahan2003,Hedges2009,Elmatad2010,Elmatad2013}; and non-equilibrium glassy states will be those where excitations would be very scarce \cite{Keys2015}.
In our highly simplified approach the stability of the stable glass (i.e.\ its fictive temperature), which in experiments is a consequence of preparation (substrate temperature, deposition rate, etc.) is encoded in the energy barrier $U$ for violation of the constraint.  This makes the bulk relaxation a nucleation and growth process that follows Avrami like scaling: excitations have to be created spontaneously in the bulk and they subsequently relax their neighbourhood in a facilitated manner (this latter process has been analysed in detail in $d>1$ East models \cite{Chleboun2014}). 

We additionally tried to replicate the same phenomenology by using a hard-constrained three-dimensional East model with a number of excitations in the bulk corresponding to a given fictive temperature much lower than the annealing temperature \cite{Keys2015}. However, the bulk relaxation of such a system, while qualitatively sigmoidal for very low fictive temperatures, does not follow the Avrami functional form. The discrepancy is especially conspicuous at the initial stages, where the growth is much faster than that predicted by the Avrami fit, and leads to a crossover considerably less well defined that that shown in Fig.~\ref{fig1} (b) or in the experimental results reported in Ref. \cite{sepulveda2014}. Results arising from our efforts in this connection indicate, though, that reducing the fictive temperature drastically, and making larger and larger systems so as to accommodate at least one or a very small number of initial bulk excitation might make the relaxation closer to an Avrami form, but the realisation that this requires experimentally unrealistically small fictive temperatures, combined with the computational difficulties involved, made us abandon this possibility. Further work may be helpful to properly elucidate whether the softness is really required to replicate the phenomenology displayed by experimental systems.

While our simplified modelling does capture the competition between front and bulk dynamics as seen experimentally, an implication is that the glass stability is encoded in a parameter rather than in the configuration reached after preparation.  This limitation is a consequence of the fact that a facilitated model is ``too coarse-grained'' with all structural features being removed from the description.  As such there is no concept of excitation confinement, which may play a significant role in highly inactive configurations (some aspects of this confinement can be seen in the difference in excitation distributions between equilibrium and inactive/non-equilibrium configurations, which can be studied in detail with large-deviation methods \cite{Keys2015}).  A more complete approach for the bulk relaxation seems to be that of the recent Ref.\ \cite{Jack2016}, that considers plaquette models (which as glass models are slightly less coarse-grained than facilitated models as they retain non-trivial structural features).  In terms of their excitation or defect variables their dynamics is effectively kinetically constrained, also with a soft constraint, as for  the soft East model we consider.  The configuration space of plaquette models, however, also allows for states where defects are confined, which is related to the existence of phase transitions of such systems in the presence of external fields or when two copies are coupled \cite{Sasa2010,Garrahan2014,Jack2016b}.   The approach of \cite{Jack2016} thus allows to encode the glass stability in the system configuration, presumably one that would be reached after preparation.  Given the similarities between certain plaquette models and East models \cite{Garrahan2002} we expect that an analysis like the the one presented here of a system with a permanently excited free surface and a stable bulk will lead to an analogous crossover between front and bulk mechanisms for ultrastable glass melting.

\begin{acknowledgments}
\noindent We thank Mark Ediger for insightful discussions. We also acknowledge financial support from EPSRC Grant no.\ EP/M014266/1. Our work has benefited from the computational resources and assistance  provided  by the University of Nottingham High Performance Computing service.
\end{acknowledgments}


\begin{thebibliography}{54}%
\makeatletter
\providecommand \@ifxundefined [1]{%
 \@ifx{#1\undefined}
}%
\providecommand \@ifnum [1]{%
 \ifnum #1\expandafter \@firstoftwo
 \else \expandafter \@secondoftwo
 \fi
}%
\providecommand \@ifx [1]{%
 \ifx #1\expandafter \@firstoftwo
 \else \expandafter \@secondoftwo
 \fi
}%
\providecommand \natexlab [1]{#1}%
\providecommand \enquote  [1]{``#1''}%
\providecommand \bibnamefont  [1]{#1}%
\providecommand \bibfnamefont [1]{#1}%
\providecommand \citenamefont [1]{#1}%
\providecommand \href@noop [0]{\@secondoftwo}%
\providecommand \href [0]{\begingroup \@sanitize@url \@href}%
\providecommand \@href[1]{\@@startlink{#1}\@@href}%
\providecommand \@@href[1]{\endgroup#1\@@endlink}%
\providecommand \@sanitize@url [0]{\catcode `\\12\catcode `\$12\catcode
  `\&12\catcode `\#12\catcode `\^12\catcode `\_12\catcode `\%12\relax}%
\providecommand \@@startlink[1]{}%
\providecommand \@@endlink[0]{}%
\providecommand \url  [0]{\begingroup\@sanitize@url \@url }%
\providecommand \@url [1]{\endgroup\@href {#1}{\urlprefix }}%
\providecommand \urlprefix  [0]{URL }%
\providecommand \Eprint [0]{\href }%
\providecommand \doibase [0]{http://dx.doi.org/}%
\providecommand \selectlanguage [0]{\@gobble}%
\providecommand \bibinfo  [0]{\@secondoftwo}%
\providecommand \bibfield  [0]{\@secondoftwo}%
\providecommand \translation [1]{[#1]}%
\providecommand \BibitemOpen [0]{}%
\providecommand \bibitemStop [0]{}%
\providecommand \bibitemNoStop [0]{.\EOS\space}%
\providecommand \EOS [0]{\spacefactor3000\relax}%
\providecommand \BibitemShut  [1]{\csname bibitem#1\endcsname}%
\let\auto@bib@innerbib\@empty
\bibitem [{\citenamefont {Struik}(1977)}]{Struik1977}%
  \BibitemOpen
  \bibfield  {author} {\bibinfo {author} {\bibfnamefont {L.~C.~E.}\
  \bibnamefont {Struik}},\ }\emph {\bibinfo {title} {Physical aging in
  amorphous polymers and other materials}},\ \href@noop {} {Ph.D. thesis},\
  \bibinfo  {school} {TU Delft} (\bibinfo {year} {1977})\BibitemShut {NoStop}%
\bibitem [{\citenamefont {Ediger}\ \emph {et~al.}(1996)\citenamefont {Ediger},
  \citenamefont {Angell},\ and\ \citenamefont {Nagel}}]{Ediger1996}%
  \BibitemOpen
  \bibfield  {author} {\bibinfo {author} {\bibfnamefont {M.~D.}\ \bibnamefont
  {Ediger}}, \bibinfo {author} {\bibfnamefont {C.~A.}\ \bibnamefont {Angell}},
  \ and\ \bibinfo {author} {\bibfnamefont {S.~R.}\ \bibnamefont {Nagel}},\
  }\href@noop {} {\bibfield  {journal} {\bibinfo  {journal} {J. Phys. Chem.}\
  }\textbf {\bibinfo {volume} {100}},\ \bibinfo {pages} {13200} (\bibinfo
  {year} {1996})}\BibitemShut {NoStop}%
\bibitem [{\citenamefont {Angell}\ \emph {et~al.}(2000)\citenamefont {Angell},
  \citenamefont {Ngai}, \citenamefont {McKenna}, \citenamefont {McMillan},\
  and\ \citenamefont {Martin}}]{Angell2000}%
  \BibitemOpen
  \bibfield  {author} {\bibinfo {author} {\bibfnamefont {C.~A.}\ \bibnamefont
  {Angell}}, \bibinfo {author} {\bibfnamefont {K.~L.}\ \bibnamefont {Ngai}},
  \bibinfo {author} {\bibfnamefont {G.~B.}\ \bibnamefont {McKenna}}, \bibinfo
  {author} {\bibfnamefont {P.~F.}\ \bibnamefont {McMillan}}, \ and\ \bibinfo
  {author} {\bibfnamefont {S.~W.}\ \bibnamefont {Martin}},\ }\href@noop {}
  {\bibfield  {journal} {\bibinfo  {journal} {J. Appl. Phys.}\ }\textbf
  {\bibinfo {volume} {88}},\ \bibinfo {pages} {3113} (\bibinfo {year}
  {2000})}\BibitemShut {NoStop}%
\bibitem [{\citenamefont {Binder}\ and\ \citenamefont
  {Kob}(2011)}]{Binder2011}%
  \BibitemOpen
  \bibfield  {author} {\bibinfo {author} {\bibfnamefont {K.}~\bibnamefont
  {Binder}}\ and\ \bibinfo {author} {\bibfnamefont {W.}~\bibnamefont {Kob}},\
  }\href@noop {} {\emph {\bibinfo {title} {Glassy Materials and Disordered
  Solids: An Introduction to Their Statistical Mechanics (Revised Edition)}}}\
  (\bibinfo  {publisher} {World Scientific},\ \bibinfo {year}
  {2011})\BibitemShut {NoStop}%
\bibitem [{\citenamefont {Swallen}\ \emph {et~al.}(2007)\citenamefont
  {Swallen}, \citenamefont {Kearns}, \citenamefont {Mapes}, \citenamefont
  {Kim}, \citenamefont {McMahon}, \citenamefont {Ediger}, \citenamefont {Wu},
  \citenamefont {Yu},\ and\ \citenamefont {Satija}}]{swallen2007}%
  \BibitemOpen
  \bibfield  {author} {\bibinfo {author} {\bibfnamefont {S.~F.}\ \bibnamefont
  {Swallen}}, \bibinfo {author} {\bibfnamefont {K.~L.}\ \bibnamefont {Kearns}},
  \bibinfo {author} {\bibfnamefont {M.~K.}\ \bibnamefont {Mapes}}, \bibinfo
  {author} {\bibfnamefont {Y.~S.}\ \bibnamefont {Kim}}, \bibinfo {author}
  {\bibfnamefont {R.~J.}\ \bibnamefont {McMahon}}, \bibinfo {author}
  {\bibfnamefont {M.~D.}\ \bibnamefont {Ediger}}, \bibinfo {author}
  {\bibfnamefont {T.}~\bibnamefont {Wu}}, \bibinfo {author} {\bibfnamefont
  {L.}~\bibnamefont {Yu}}, \ and\ \bibinfo {author} {\bibfnamefont
  {S.}~\bibnamefont {Satija}},\ }\href@noop {} {\bibfield  {journal} {\bibinfo
  {journal} {Science}\ }\textbf {\bibinfo {volume} {315}},\ \bibinfo {pages}
  {353} (\bibinfo {year} {2007})}\BibitemShut {NoStop}%
\bibitem [{\citenamefont {Kearns}\ \emph {et~al.}(2010)\citenamefont {Kearns},
  \citenamefont {Ediger}, \citenamefont {Huth},\ and\ \citenamefont
  {Schick}}]{kearns2009}%
  \BibitemOpen
  \bibfield  {author} {\bibinfo {author} {\bibfnamefont {K.~L.}\ \bibnamefont
  {Kearns}}, \bibinfo {author} {\bibfnamefont {M.~D.}\ \bibnamefont {Ediger}},
  \bibinfo {author} {\bibfnamefont {H.}~\bibnamefont {Huth}}, \ and\ \bibinfo
  {author} {\bibfnamefont {C.}~\bibnamefont {Schick}},\ }\href@noop {}
  {\bibfield  {journal} {\bibinfo  {journal} {J. Phys. Chem. Lett.}\ }\textbf
  {\bibinfo {volume} {1}},\ \bibinfo {pages} {388} (\bibinfo {year}
  {2010})}\BibitemShut {NoStop}%
\bibitem [{\citenamefont {Swallen}\ \emph {et~al.}(2010)\citenamefont
  {Swallen}, \citenamefont {Windsor}, \citenamefont {McMahon}, \citenamefont
  {Ediger},\ and\ \citenamefont {Mates}}]{swallen2010}%
  \BibitemOpen
  \bibfield  {author} {\bibinfo {author} {\bibfnamefont {S.~F.}\ \bibnamefont
  {Swallen}}, \bibinfo {author} {\bibfnamefont {K.}~\bibnamefont {Windsor}},
  \bibinfo {author} {\bibfnamefont {R.~J.}\ \bibnamefont {McMahon}}, \bibinfo
  {author} {\bibfnamefont {M.~D.}\ \bibnamefont {Ediger}}, \ and\ \bibinfo
  {author} {\bibfnamefont {T.~E.}\ \bibnamefont {Mates}},\ }\href@noop {}
  {\bibfield  {journal} {\bibinfo  {journal} {J. Phys. Chem. B}\ }\textbf
  {\bibinfo {volume} {114}},\ \bibinfo {pages} {2635} (\bibinfo {year}
  {2010})}\BibitemShut {NoStop}%
\bibitem [{\citenamefont {Dalal}\ \emph {et~al.}(2012)\citenamefont {Dalal},
  \citenamefont {Sep{\'u}lveda}, \citenamefont {Pribil}, \citenamefont
  {Fakhraai},\ and\ \citenamefont {Ediger}}]{dalal2012}%
  \BibitemOpen
  \bibfield  {author} {\bibinfo {author} {\bibfnamefont {S.~S.}\ \bibnamefont
  {Dalal}}, \bibinfo {author} {\bibfnamefont {A.}~\bibnamefont
  {Sep{\'u}lveda}}, \bibinfo {author} {\bibfnamefont {G.~K.}\ \bibnamefont
  {Pribil}}, \bibinfo {author} {\bibfnamefont {Z.}~\bibnamefont {Fakhraai}}, \
  and\ \bibinfo {author} {\bibfnamefont {M.~D.}\ \bibnamefont {Ediger}},\
  }\href@noop {} {\bibfield  {journal} {\bibinfo  {journal} {J. Chem. Phys.}\
  }\textbf {\bibinfo {volume} {136}},\ \bibinfo {pages} {204501} (\bibinfo
  {year} {2012})}\BibitemShut {NoStop}%
\bibitem [{\citenamefont {Sep{\'u}lveda}\ \emph {et~al.}(2013)\citenamefont
  {Sep{\'u}lveda}, \citenamefont {Swallen},\ and\ \citenamefont
  {Ediger}}]{sepulveda2013}%
  \BibitemOpen
  \bibfield  {author} {\bibinfo {author} {\bibfnamefont {A.}~\bibnamefont
  {Sep{\'u}lveda}}, \bibinfo {author} {\bibfnamefont {S.~F.}\ \bibnamefont
  {Swallen}}, \ and\ \bibinfo {author} {\bibfnamefont {M.~D.}\ \bibnamefont
  {Ediger}},\ }\href@noop {} {\bibfield  {journal} {\bibinfo  {journal} {J.
  Chem. Phys.}\ }\textbf {\bibinfo {volume} {138}},\ \bibinfo {pages} {12A517}
  (\bibinfo {year} {2013})}\BibitemShut {NoStop}%
\bibitem [{\citenamefont {Swallen}\ \emph {et~al.}(2009)\citenamefont
  {Swallen}, \citenamefont {Traynor}, \citenamefont {McMahon}, \citenamefont
  {Ediger},\ and\ \citenamefont {Mates}}]{swallen2009}%
  \BibitemOpen
  \bibfield  {author} {\bibinfo {author} {\bibfnamefont {S.~F.}\ \bibnamefont
  {Swallen}}, \bibinfo {author} {\bibfnamefont {K.}~\bibnamefont {Traynor}},
  \bibinfo {author} {\bibfnamefont {R.~J.}\ \bibnamefont {McMahon}}, \bibinfo
  {author} {\bibfnamefont {M.~D.}\ \bibnamefont {Ediger}}, \ and\ \bibinfo
  {author} {\bibfnamefont {T.~E.}\ \bibnamefont {Mates}},\ }\href@noop {}
  {\bibfield  {journal} {\bibinfo  {journal} {Phys. Rev. Lett.}\ }\textbf
  {\bibinfo {volume} {102}},\ \bibinfo {pages} {065503} (\bibinfo {year}
  {2009})}\BibitemShut {NoStop}%
\bibitem [{\citenamefont {Rodr{\'\i}guez-Tinoco}\ \emph
  {et~al.}(2014)\citenamefont {Rodr{\'\i}guez-Tinoco}, \citenamefont
  {Gonzalez-Silveira}, \citenamefont {R{\`a}fols-Rib{\'e}}, \citenamefont
  {Lopeand{\'\i}a}, \citenamefont {Clavaguera-Mora},\ and\ \citenamefont
  {Rodr{\'\i}guez-Viejo}}]{rodriguez2014}%
  \BibitemOpen
  \bibfield  {author} {\bibinfo {author} {\bibfnamefont {C.}~\bibnamefont
  {Rodr{\'\i}guez-Tinoco}}, \bibinfo {author} {\bibfnamefont {M.}~\bibnamefont
  {Gonzalez-Silveira}}, \bibinfo {author} {\bibfnamefont {J.}~\bibnamefont
  {R{\`a}fols-Rib{\'e}}}, \bibinfo {author} {\bibfnamefont {A.~F.}\
  \bibnamefont {Lopeand{\'\i}a}}, \bibinfo {author} {\bibfnamefont {M.~T.}\
  \bibnamefont {Clavaguera-Mora}}, \ and\ \bibinfo {author} {\bibfnamefont
  {J.}~\bibnamefont {Rodr{\'\i}guez-Viejo}},\ }\href@noop {} {\bibfield
  {journal} {\bibinfo  {journal} {J. Phys. Chem. B}\ }\textbf {\bibinfo
  {volume} {118}},\ \bibinfo {pages} {10795} (\bibinfo {year}
  {2014})}\BibitemShut {NoStop}%
\bibitem [{\citenamefont {Rodr{\'\i}guez-Tinoco}\ \emph
  {et~al.}(2015)\citenamefont {Rodr{\'\i}guez-Tinoco}, \citenamefont
  {Gonzalez-Silveira}, \citenamefont {R{\`a}fols-Rib{\'e}}, \citenamefont
  {Lopeand{\'\i}a},\ and\ \citenamefont
  {Rodr{\'\i}guez-Viejo}}]{rodriguez2015}%
  \BibitemOpen
  \bibfield  {author} {\bibinfo {author} {\bibfnamefont {C.}~\bibnamefont
  {Rodr{\'\i}guez-Tinoco}}, \bibinfo {author} {\bibfnamefont {M.}~\bibnamefont
  {Gonzalez-Silveira}}, \bibinfo {author} {\bibfnamefont {J.}~\bibnamefont
  {R{\`a}fols-Rib{\'e}}}, \bibinfo {author} {\bibfnamefont {A.~F.}\
  \bibnamefont {Lopeand{\'\i}a}}, \ and\ \bibinfo {author} {\bibfnamefont
  {J.}~\bibnamefont {Rodr{\'\i}guez-Viejo}},\ }\href@noop {} {\bibfield
  {journal} {\bibinfo  {journal} {Phys. Chem. Chem. Phys.}\ }\textbf {\bibinfo
  {volume} {17}},\ \bibinfo {pages} {31195} (\bibinfo {year}
  {2015})}\BibitemShut {NoStop}%
\bibitem [{\citenamefont {Tylinski}\ \emph {et~al.}(2015)\citenamefont
  {Tylinski}, \citenamefont {Sep{\'u}lveda}, \citenamefont {Walters},
  \citenamefont {Chua}, \citenamefont {Schick},\ and\ \citenamefont
  {Ediger}}]{tylinski2015}%
  \BibitemOpen
  \bibfield  {author} {\bibinfo {author} {\bibfnamefont {M.}~\bibnamefont
  {Tylinski}}, \bibinfo {author} {\bibfnamefont {A.}~\bibnamefont
  {Sep{\'u}lveda}}, \bibinfo {author} {\bibfnamefont {D.~M.}\ \bibnamefont
  {Walters}}, \bibinfo {author} {\bibfnamefont {Y.~Z.}\ \bibnamefont {Chua}},
  \bibinfo {author} {\bibfnamefont {C.}~\bibnamefont {Schick}}, \ and\ \bibinfo
  {author} {\bibfnamefont {M.~D.}\ \bibnamefont {Ediger}},\ }\href@noop {}
  {\bibfield  {journal} {\bibinfo  {journal} {J. Chem. Phys.}\ }\textbf
  {\bibinfo {volume} {143}},\ \bibinfo {pages} {244509} (\bibinfo {year}
  {2015})}\BibitemShut {NoStop}%
\bibitem [{\citenamefont {Zhang}\ \emph {et~al.}(2016)\citenamefont {Zhang},
  \citenamefont {Glor}, \citenamefont {Li}, \citenamefont {Liu}, \citenamefont
  {Wahid}, \citenamefont {Zhang}, \citenamefont {Riggleman}, \citenamefont
  {Fakhraai},\ and\ \citenamefont {Fakhraai}}]{zhang2016}%
  \BibitemOpen
  \bibfield  {author} {\bibinfo {author} {\bibfnamefont {Y.}~\bibnamefont
  {Zhang}}, \bibinfo {author} {\bibfnamefont {E.}~\bibnamefont {Glor}},
  \bibinfo {author} {\bibfnamefont {M.}~\bibnamefont {Li}}, \bibinfo {author}
  {\bibfnamefont {T.}~\bibnamefont {Liu}}, \bibinfo {author} {\bibfnamefont
  {K.}~\bibnamefont {Wahid}}, \bibinfo {author} {\bibfnamefont
  {W.}~\bibnamefont {Zhang}}, \bibinfo {author} {\bibfnamefont
  {R.}~\bibnamefont {Riggleman}}, \bibinfo {author} {\bibfnamefont
  {Z.}~\bibnamefont {Fakhraai}}, \ and\ \bibinfo {author} {\bibfnamefont
  {Z.}~\bibnamefont {Fakhraai}},\ }\href@noop {} {\bibfield  {journal}
  {\bibinfo  {journal} {arXiv preprint arXiv:1603.07244}\ } (\bibinfo {year}
  {2016})}\BibitemShut {NoStop}%
\bibitem [{\citenamefont {Sep{\'u}lveda}\ \emph {et~al.}(2014)\citenamefont
  {Sep{\'u}lveda}, \citenamefont {Tylinski}, \citenamefont {Guiseppi-Elie},
  \citenamefont {Richert},\ and\ \citenamefont {Ediger}}]{sepulveda2014}%
  \BibitemOpen
  \bibfield  {author} {\bibinfo {author} {\bibfnamefont {A.}~\bibnamefont
  {Sep{\'u}lveda}}, \bibinfo {author} {\bibfnamefont {M.}~\bibnamefont
  {Tylinski}}, \bibinfo {author} {\bibfnamefont {A.}~\bibnamefont
  {Guiseppi-Elie}}, \bibinfo {author} {\bibfnamefont {R.}~\bibnamefont
  {Richert}}, \ and\ \bibinfo {author} {\bibfnamefont {M.~D.}\ \bibnamefont
  {Ediger}},\ }\href@noop {} {\bibfield  {journal} {\bibinfo  {journal} {Phys.
  Rev. Lett.}\ }\textbf {\bibinfo {volume} {113}},\ \bibinfo {pages} {045901}
  (\bibinfo {year} {2014})}\BibitemShut {NoStop}%
\bibitem [{\citenamefont {Chandler}\ and\ \citenamefont
  {Garrahan}(2010)}]{Chandler2010}%
  \BibitemOpen
  \bibfield  {author} {\bibinfo {author} {\bibfnamefont {D.}~\bibnamefont
  {Chandler}}\ and\ \bibinfo {author} {\bibfnamefont {J.~P.}\ \bibnamefont
  {Garrahan}},\ }\href
  {http://dx.doi.org/10.1146/annurev.physchem.040808.090405} {\bibfield
  {journal} {\bibinfo  {journal} {Annu. Rev. Phys. Chem.}\ }\textbf {\bibinfo
  {volume} {61}},\ \bibinfo {pages} {191} (\bibinfo {year} {2010})}\BibitemShut
  {NoStop}%
\bibitem [{\citenamefont {Ritort}\ and\ \citenamefont
  {Sollich}(2003)}]{Ritort2003}%
  \BibitemOpen
  \bibfield  {author} {\bibinfo {author} {\bibfnamefont {F.}~\bibnamefont
  {Ritort}}\ and\ \bibinfo {author} {\bibfnamefont {P.}~\bibnamefont
  {Sollich}},\ }\href@noop {} {\bibfield  {journal} {\bibinfo  {journal} {Adv.
  Phys.}\ }\textbf {\bibinfo {volume} {52}},\ \bibinfo {pages} {219} (\bibinfo
  {year} {2003})}\BibitemShut {NoStop}%
\bibitem [{\citenamefont {Ashton}\ \emph {et~al.}(2005)\citenamefont {Ashton},
  \citenamefont {Hedges},\ and\ \citenamefont {Garrahan}}]{Ashton2005}%
  \BibitemOpen
  \bibfield  {author} {\bibinfo {author} {\bibfnamefont {D.~J.}\ \bibnamefont
  {Ashton}}, \bibinfo {author} {\bibfnamefont {L.~O.}\ \bibnamefont {Hedges}},
  \ and\ \bibinfo {author} {\bibfnamefont {J.~P.}\ \bibnamefont {Garrahan}},\
  }\href {http://stacks.iop.org/1742-5468/2005/i=12/a=P12010} {\bibfield
  {journal} {\bibinfo  {journal} {J. Stat. Mech.}\ }\textbf {\bibinfo {volume}
  {2005}},\ \bibinfo {pages} {P12010} (\bibinfo {year} {2005})}\BibitemShut
  {NoStop}%
\bibitem [{\citenamefont {Berthier}\ and\ \citenamefont
  {Garrahan}(2005)}]{Berthier2005}%
  \BibitemOpen
  \bibfield  {author} {\bibinfo {author} {\bibfnamefont {L.}~\bibnamefont
  {Berthier}}\ and\ \bibinfo {author} {\bibfnamefont {J.~P.}\ \bibnamefont
  {Garrahan}},\ }\href@noop {} {\bibfield  {journal} {\bibinfo  {journal} {J.
  Phys. Chem. B}\ }\textbf {\bibinfo {volume} {109}},\ \bibinfo {pages} {3578}
  (\bibinfo {year} {2005})}\BibitemShut {NoStop}%
\bibitem [{\citenamefont {Chleboun}\ \emph
  {et~al.}(2014{\natexlab{a}})\citenamefont {Chleboun}, \citenamefont
  {Faggionato},\ and\ \citenamefont {Martinelli}}]{Chleboun2014}%
  \BibitemOpen
  \bibfield  {author} {\bibinfo {author} {\bibfnamefont {P.}~\bibnamefont
  {Chleboun}}, \bibinfo {author} {\bibfnamefont {A.}~\bibnamefont
  {Faggionato}}, \ and\ \bibinfo {author} {\bibfnamefont {F.}~\bibnamefont
  {Martinelli}},\ }\href@noop {} {\bibfield  {journal} {\bibinfo  {journal}
  {EPL}\ }\textbf {\bibinfo {volume} {107}},\ \bibinfo {pages} {36002}
  (\bibinfo {year} {2014}{\natexlab{a}})}\BibitemShut {NoStop}%
\bibitem [{\citenamefont {Chleboun}\ \emph
  {et~al.}(2014{\natexlab{b}})\citenamefont {Chleboun}, \citenamefont
  {Faggionato},\ and\ \citenamefont {Martinelli}}]{Chleboun2014b}%
  \BibitemOpen
  \bibfield  {author} {\bibinfo {author} {\bibfnamefont {P.}~\bibnamefont
  {Chleboun}}, \bibinfo {author} {\bibfnamefont {A.}~\bibnamefont
  {Faggionato}}, \ and\ \bibinfo {author} {\bibfnamefont {F.}~\bibnamefont
  {Martinelli}},\ }\href@noop {} {\bibfield  {journal} {\bibinfo  {journal}
  {arXiv:1404.7257}\ } (\bibinfo {year} {2014}{\natexlab{b}})}\BibitemShut
  {NoStop}%
\bibitem [{\citenamefont {Chleboun}\ \emph {et~al.}(2015)\citenamefont
  {Chleboun}, \citenamefont {Faggionato},\ and\ \citenamefont
  {Martinelli}}]{Chleboun2015}%
  \BibitemOpen
  \bibfield  {author} {\bibinfo {author} {\bibfnamefont {P.}~\bibnamefont
  {Chleboun}}, \bibinfo {author} {\bibfnamefont {A.}~\bibnamefont
  {Faggionato}}, \ and\ \bibinfo {author} {\bibfnamefont {F.}~\bibnamefont
  {Martinelli}},\ }\href@noop {} {\bibfield  {journal} {\bibinfo  {journal}
  {arXiv:1501.02240}\ } (\bibinfo {year} {2015})}\BibitemShut {NoStop}%
\bibitem [{\citenamefont {Elmatad}\ and\ \citenamefont
  {Jack}(2013)}]{Elmatad2013}%
  \BibitemOpen
  \bibfield  {author} {\bibinfo {author} {\bibfnamefont {Y.~S.}\ \bibnamefont
  {Elmatad}}\ and\ \bibinfo {author} {\bibfnamefont {R.~L.}\ \bibnamefont
  {Jack}},\ }\href@noop {} {\bibfield  {journal} {\bibinfo  {journal} {J. Chem.
  Phys.}\ }\textbf {\bibinfo {volume} {138}},\ \bibinfo {pages} {12A531}
  (\bibinfo {year} {2013})}\BibitemShut {NoStop}%
\bibitem [{\citenamefont {Garrahan}(2002)}]{Garrahan2002}%
  \BibitemOpen
  \bibfield  {author} {\bibinfo {author} {\bibfnamefont {J.~P.}\ \bibnamefont
  {Garrahan}},\ }\href {\doibase PII S0953-8984(02)31606-0} {\bibfield
  {journal} {\bibinfo  {journal} {J. Phys. Condens. Matter}\ }\textbf {\bibinfo
  {volume} {14}},\ \bibinfo {pages} {1571} (\bibinfo {year}
  {2002})}\BibitemShut {NoStop}%
\bibitem [{\citenamefont {Wolynes}(2009)}]{Wolynes2009}%
  \BibitemOpen
  \bibfield  {author} {\bibinfo {author} {\bibfnamefont {P.~G.}\ \bibnamefont
  {Wolynes}},\ }\href@noop {} {\bibfield  {journal} {\bibinfo  {journal} {Proc.
  Natl. Acad. Sci. USA}\ }\textbf {\bibinfo {volume} {106}},\ \bibinfo {pages}
  {1353} (\bibinfo {year} {2009})}\BibitemShut {NoStop}%
\bibitem [{\citenamefont {Singh}\ and\ \citenamefont
  {de~Pablo}(2011)}]{Singh2011}%
  \BibitemOpen
  \bibfield  {author} {\bibinfo {author} {\bibfnamefont {S.}~\bibnamefont
  {Singh}}\ and\ \bibinfo {author} {\bibfnamefont {J.~J.}\ \bibnamefont
  {de~Pablo}},\ }\href@noop {} {\bibfield  {journal} {\bibinfo  {journal} {J.
  Chem. Phys.}\ }\textbf {\bibinfo {volume} {134}},\ \bibinfo {pages} {194903}
  (\bibinfo {year} {2011})}\BibitemShut {NoStop}%
\bibitem [{\citenamefont {Wisitsorasak}\ and\ \citenamefont
  {Wolynes}(2013)}]{Wisitsorasak2013}%
  \BibitemOpen
  \bibfield  {author} {\bibinfo {author} {\bibfnamefont {A.}~\bibnamefont
  {Wisitsorasak}}\ and\ \bibinfo {author} {\bibfnamefont {P.}~\bibnamefont
  {Wolynes}},\ }\href@noop {} {\bibfield  {journal} {\bibinfo  {journal} {Phys.
  Rev. E}\ }\textbf {\bibinfo {volume} {88}},\ \bibinfo {pages} {022308}
  (\bibinfo {year} {2013})}\BibitemShut {NoStop}%
\bibitem [{\citenamefont {Mirigian}\ and\ \citenamefont
  {Schweizer}(2014)}]{Mirigian2014}%
  \BibitemOpen
  \bibfield  {author} {\bibinfo {author} {\bibfnamefont {S.}~\bibnamefont
  {Mirigian}}\ and\ \bibinfo {author} {\bibfnamefont {K.~S.}\ \bibnamefont
  {Schweizer}},\ }\href@noop {} {\bibfield  {journal} {\bibinfo  {journal} {J.
  Chem. Phys.}\ }\textbf {\bibinfo {volume} {141}},\ \bibinfo {pages} {161103}
  (\bibinfo {year} {2014})}\BibitemShut {NoStop}%
\bibitem [{\citenamefont {L{\'e}onard}\ and\ \citenamefont
  {Harrowell}(2010)}]{leonard2010}%
  \BibitemOpen
  \bibfield  {author} {\bibinfo {author} {\bibfnamefont {S.}~\bibnamefont
  {L{\'e}onard}}\ and\ \bibinfo {author} {\bibfnamefont {P.}~\bibnamefont
  {Harrowell}},\ }\href@noop {} {\bibfield  {journal} {\bibinfo  {journal} {J.
  Chem. Phys.}\ }\textbf {\bibinfo {volume} {133}},\ \bibinfo {pages} {244502}
  (\bibinfo {year} {2010})}\BibitemShut {NoStop}%
\bibitem [{\citenamefont {Douglass}\ and\ \citenamefont
  {Harrowell}(2013)}]{douglass2013}%
  \BibitemOpen
  \bibfield  {author} {\bibinfo {author} {\bibfnamefont {I.}~\bibnamefont
  {Douglass}}\ and\ \bibinfo {author} {\bibfnamefont {P.}~\bibnamefont
  {Harrowell}},\ }\href@noop {} {\bibfield  {journal} {\bibinfo  {journal} {J.
  Chem. Phys.}\ }\textbf {\bibinfo {volume} {138}},\ \bibinfo {pages} {12A516}
  (\bibinfo {year} {2013})}\BibitemShut {NoStop}%
\bibitem [{\citenamefont {Jack}\ and\ \citenamefont
  {Berthier}(2016)}]{Jack2016}%
  \BibitemOpen
  \bibfield  {author} {\bibinfo {author} {\bibfnamefont {R.~L.}\ \bibnamefont
  {Jack}}\ and\ \bibinfo {author} {\bibfnamefont {L.}~\bibnamefont
  {Berthier}},\ }\href@noop {} {\bibfield  {journal} {\bibinfo  {journal}
  {arXiv preprint arXiv:1603.05017}\ } (\bibinfo {year} {2016})}\BibitemShut
  {NoStop}%
\bibitem [{\citenamefont {Sollich}\ and\ \citenamefont
  {Evans}(1999)}]{Sollich1999}%
  \BibitemOpen
  \bibfield  {author} {\bibinfo {author} {\bibfnamefont {P.}~\bibnamefont
  {Sollich}}\ and\ \bibinfo {author} {\bibfnamefont {M.~R.}\ \bibnamefont
  {Evans}},\ }\href@noop {} {\bibfield  {journal} {\bibinfo  {journal} {Phys.
  Rev. Lett.}\ }\textbf {\bibinfo {volume} {83}},\ \bibinfo {pages} {3238}
  (\bibinfo {year} {1999})}\BibitemShut {NoStop}%
\bibitem [{\citenamefont {Elmatad}\ \emph {et~al.}(2009)\citenamefont
  {Elmatad}, \citenamefont {Chandler},\ and\ \citenamefont
  {Garrahan}}]{Elmatad2009}%
  \BibitemOpen
  \bibfield  {author} {\bibinfo {author} {\bibfnamefont {Y.~S.}\ \bibnamefont
  {Elmatad}}, \bibinfo {author} {\bibfnamefont {D.}~\bibnamefont {Chandler}}, \
  and\ \bibinfo {author} {\bibfnamefont {J.~P.}\ \bibnamefont {Garrahan}},\
  }\href {http://pubs.acs.org/doi/abs/10.1021/jp810362g} {\bibfield  {journal}
  {\bibinfo  {journal} {J. Phys. Chem. B}\ }\textbf {\bibinfo {volume} {113}},\
  \bibinfo {pages} {5563} (\bibinfo {year} {2009})}\BibitemShut {NoStop}%
\bibitem [{\citenamefont {Garrahan}\ and\ \citenamefont
  {Chandler}(2002)}]{Garrahan2002a}%
  \BibitemOpen
  \bibfield  {author} {\bibinfo {author} {\bibfnamefont {J.~P.}\ \bibnamefont
  {Garrahan}}\ and\ \bibinfo {author} {\bibfnamefont {D.}~\bibnamefont
  {Chandler}},\ }\href@noop {} {\bibfield  {journal} {\bibinfo  {journal}
  {Phys. Rev. Lett.}\ }\textbf {\bibinfo {volume} {89}},\ \bibinfo {pages}
  {035704} (\bibinfo {year} {2002})}\BibitemShut {NoStop}%
\bibitem [{\citenamefont {Jung}\ \emph {et~al.}(2004)\citenamefont {Jung},
  \citenamefont {Garrahan},\ and\ \citenamefont {Chandler}}]{Jung2004}%
  \BibitemOpen
  \bibfield  {author} {\bibinfo {author} {\bibfnamefont {Y.}~\bibnamefont
  {Jung}}, \bibinfo {author} {\bibfnamefont {J.}~\bibnamefont {Garrahan}}, \
  and\ \bibinfo {author} {\bibfnamefont {D.}~\bibnamefont {Chandler}},\
  }\href@noop {} {\bibfield  {journal} {\bibinfo  {journal} {Phys. Rev. E}\
  }\textbf {\bibinfo {volume} {69}},\ \bibinfo {pages} {061205} (\bibinfo
  {year} {2004})}\BibitemShut {NoStop}%
\bibitem [{\citenamefont {Blondel}\ and\ \citenamefont
  {Toninelli}(2014)}]{Blondel2014}%
  \BibitemOpen
  \bibfield  {author} {\bibinfo {author} {\bibfnamefont {O.}~\bibnamefont
  {Blondel}}\ and\ \bibinfo {author} {\bibfnamefont {C.}~\bibnamefont
  {Toninelli}},\ }\href@noop {} {\bibfield  {journal} {\bibinfo  {journal}
  {EPL}\ }\textbf {\bibinfo {volume} {107}},\ \bibinfo {pages} {26005}
  (\bibinfo {year} {2014})}\BibitemShut {NoStop}%
\bibitem [{\citenamefont {Keys}\ \emph {et~al.}(2013)\citenamefont {Keys},
  \citenamefont {Garrahan},\ and\ \citenamefont {Chandler}}]{Keys2013}%
  \BibitemOpen
  \bibfield  {author} {\bibinfo {author} {\bibfnamefont {A.~S.}\ \bibnamefont
  {Keys}}, \bibinfo {author} {\bibfnamefont {J.~P.}\ \bibnamefont {Garrahan}},
  \ and\ \bibinfo {author} {\bibfnamefont {D.}~\bibnamefont {Chandler}},\
  }\href@noop {} {\bibfield  {journal} {\bibinfo  {journal} {Proc. Natl. Acad.
  Sci. USA}\ }\textbf {\bibinfo {volume} {110}},\ \bibinfo {pages} {4482}
  (\bibinfo {year} {2013})}\BibitemShut {NoStop}%
\bibitem [{\citenamefont {Faggionato}\ \emph {et~al.}(2012)\citenamefont
  {Faggionato}, \citenamefont {Martinelli}, \citenamefont {Roberto},\ and\
  \citenamefont {Toninelli}}]{Faggionato2012}%
  \BibitemOpen
  \bibfield  {author} {\bibinfo {author} {\bibfnamefont {A.}~\bibnamefont
  {Faggionato}}, \bibinfo {author} {\bibfnamefont {F.}~\bibnamefont
  {Martinelli}}, \bibinfo {author} {\bibfnamefont {C.}~\bibnamefont {Roberto}},
  \ and\ \bibinfo {author} {\bibfnamefont {C.}~\bibnamefont {Toninelli}},\
  }\href@noop {} {\bibfield  {journal} {\bibinfo  {journal} {arXiv:1205.1607}\
  } (\bibinfo {year} {2012})}\BibitemShut {NoStop}%
\bibitem [{\citenamefont {Blondel}(2013)}]{Blondel2013}%
  \BibitemOpen
  \bibfield  {author} {\bibinfo {author} {\bibfnamefont {O.}~\bibnamefont
  {Blondel}},\ }\href@noop {} {\bibfield  {journal} {\bibinfo  {journal}
  {Stoch. Proc. Appl.}\ }\textbf {\bibinfo {volume} {123}},\ \bibinfo {pages}
  {3430} (\bibinfo {year} {2013})}\BibitemShut {NoStop}%
\bibitem [{\citenamefont {Chleboun}\ \emph
  {et~al.}(2014{\natexlab{c}})\citenamefont {Chleboun}, \citenamefont
  {Faggionato},\ and\ \citenamefont {Martinelli}}]{Chleboun2014c}%
  \BibitemOpen
  \bibfield  {author} {\bibinfo {author} {\bibfnamefont {P.}~\bibnamefont
  {Chleboun}}, \bibinfo {author} {\bibfnamefont {A.}~\bibnamefont
  {Faggionato}}, \ and\ \bibinfo {author} {\bibfnamefont {F.}~\bibnamefont
  {Martinelli}},\ }\href@noop {} {\bibfield  {journal} {\bibinfo  {journal}
  {Comm. Math. Phys.}\ }\textbf {\bibinfo {volume} {328}},\ \bibinfo {pages}
  {955} (\bibinfo {year} {2014}{\natexlab{c}})}\BibitemShut {NoStop}%
\bibitem [{\citenamefont {Kolmogorov}(1937)}]{kolmogorov1937}%
  \BibitemOpen
  \bibfield  {author} {\bibinfo {author} {\bibfnamefont {A.~E.}\ \bibnamefont
  {Kolmogorov}},\ }\href@noop {} {\bibfield  {journal} {\bibinfo  {journal}
  {Izv. Akad. Nauk. SSSR Ser. Mat.}\ }\textbf {\bibinfo {volume} {1}},\
  \bibinfo {pages} {355} (\bibinfo {year} {1937})}\BibitemShut {NoStop}%
\bibitem [{\citenamefont {Johnson}\ and\ \citenamefont
  {Mehl}(1939)}]{johnson1939}%
  \BibitemOpen
  \bibfield  {author} {\bibinfo {author} {\bibfnamefont {W.~A.}\ \bibnamefont
  {Johnson}}\ and\ \bibinfo {author} {\bibfnamefont {R.~F.}\ \bibnamefont
  {Mehl}},\ }\href@noop {} {\bibfield  {journal} {\bibinfo  {journal} {Trans.
  Am. Inst. Min. Engrs.}\ }\textbf {\bibinfo {volume} {135}},\ \bibinfo {pages}
  {396} (\bibinfo {year} {1939})}\BibitemShut {NoStop}%
\bibitem [{\citenamefont {Avrami}(1939)}]{avrami1939}%
  \BibitemOpen
  \bibfield  {author} {\bibinfo {author} {\bibfnamefont {M.}~\bibnamefont
  {Avrami}},\ }\href@noop {} {\bibfield  {journal} {\bibinfo  {journal} {J.
  Chem. Phys.}\ }\textbf {\bibinfo {volume} {7}},\ \bibinfo {pages} {1103}
  (\bibinfo {year} {1939})}\BibitemShut {NoStop}%
\bibitem [{\citenamefont {Avrami}(1940)}]{avrami1940}%
  \BibitemOpen
  \bibfield  {author} {\bibinfo {author} {\bibfnamefont {M.}~\bibnamefont
  {Avrami}},\ }\href@noop {} {\bibfield  {journal} {\bibinfo  {journal} {J.
  Chem. Phys.}\ }\textbf {\bibinfo {volume} {8}},\ \bibinfo {pages} {212}
  (\bibinfo {year} {1940})}\BibitemShut {NoStop}%
\bibitem [{\citenamefont {Avrami}(1941)}]{avrami1941}%
  \BibitemOpen
  \bibfield  {author} {\bibinfo {author} {\bibfnamefont {M.}~\bibnamefont
  {Avrami}},\ }\href@noop {} {\bibfield  {journal} {\bibinfo  {journal} {J.
  Chem. Phys.}\ }\textbf {\bibinfo {volume} {9}},\ \bibinfo {pages} {177}
  (\bibinfo {year} {1941})}\BibitemShut {NoStop}%
\bibitem [{\citenamefont {Hocky}\ \emph {et~al.}(2014)\citenamefont {Hocky},
  \citenamefont {Berthier},\ and\ \citenamefont {Reichman}}]{hocky2014}%
  \BibitemOpen
  \bibfield  {author} {\bibinfo {author} {\bibfnamefont {G.~M.}\ \bibnamefont
  {Hocky}}, \bibinfo {author} {\bibfnamefont {L.}~\bibnamefont {Berthier}}, \
  and\ \bibinfo {author} {\bibfnamefont {D.~R.}\ \bibnamefont {Reichman}},\
  }\href@noop {} {\bibfield  {journal} {\bibinfo  {journal} {J. Chem. Phys.}\
  }\textbf {\bibinfo {volume} {141}},\ \bibinfo {pages} {224503} (\bibinfo
  {year} {2014})}\BibitemShut {NoStop}%
\bibitem [{\citenamefont {Christian}(2002)}]{christian2002}%
  \BibitemOpen
  \bibfield  {author} {\bibinfo {author} {\bibfnamefont {J.~W.}\ \bibnamefont
  {Christian}},\ }\href@noop {} {\emph {\bibinfo {title} {The Theory of
  Transformations in Metals and Alloys}}}\ (\bibinfo  {publisher} {Pergamon,
  Oxford},\ \bibinfo {year} {2002})\BibitemShut {NoStop}%
\bibitem [{\citenamefont {Garrahan}\ and\ \citenamefont
  {Chandler}(2003)}]{Garrahan2003}%
  \BibitemOpen
  \bibfield  {author} {\bibinfo {author} {\bibfnamefont {J.~P.}\ \bibnamefont
  {Garrahan}}\ and\ \bibinfo {author} {\bibfnamefont {D.}~\bibnamefont
  {Chandler}},\ }\href {\doibase DOI 10.1073/pnas.1233719100} {\bibfield
  {journal} {\bibinfo  {journal} {Proc. Natl. Acad. Sci. USA}\ }\textbf
  {\bibinfo {volume} {100}},\ \bibinfo {pages} {9710} (\bibinfo {year}
  {2003})}\BibitemShut {NoStop}%
\bibitem [{\citenamefont {Hedges}\ \emph {et~al.}(2009)\citenamefont {Hedges},
  \citenamefont {Jack}, \citenamefont {Garrahan},\ and\ \citenamefont
  {Chandler}}]{Hedges2009}%
  \BibitemOpen
  \bibfield  {author} {\bibinfo {author} {\bibfnamefont {L.~O.}\ \bibnamefont
  {Hedges}}, \bibinfo {author} {\bibfnamefont {R.~L.}\ \bibnamefont {Jack}},
  \bibinfo {author} {\bibfnamefont {J.~P.}\ \bibnamefont {Garrahan}}, \ and\
  \bibinfo {author} {\bibfnamefont {D.}~\bibnamefont {Chandler}},\ }\href@noop
  {} {\bibfield  {journal} {\bibinfo  {journal} {Science}\ }\textbf {\bibinfo
  {volume} {323}},\ \bibinfo {pages} {1309} (\bibinfo {year}
  {2009})}\BibitemShut {NoStop}%
\bibitem [{\citenamefont {Elmatad}\ \emph {et~al.}(2010)\citenamefont
  {Elmatad}, \citenamefont {Jack}, \citenamefont {Chandler},\ and\
  \citenamefont {Garrahan}}]{Elmatad2010}%
  \BibitemOpen
  \bibfield  {author} {\bibinfo {author} {\bibfnamefont {Y.~S.}\ \bibnamefont
  {Elmatad}}, \bibinfo {author} {\bibfnamefont {R.~L.}\ \bibnamefont {Jack}},
  \bibinfo {author} {\bibfnamefont {D.}~\bibnamefont {Chandler}}, \ and\
  \bibinfo {author} {\bibfnamefont {J.~P.}\ \bibnamefont {Garrahan}},\
  }\href@noop {} {\bibfield  {journal} {\bibinfo  {journal} {Proc. Natl. Acad.
  Sci. USA}\ }\textbf {\bibinfo {volume} {107}},\ \bibinfo {pages} {12793}
  (\bibinfo {year} {2010})}\BibitemShut {NoStop}%
\bibitem [{\citenamefont {Keys}\ \emph {et~al.}(2015)\citenamefont {Keys},
  \citenamefont {Chandler},\ and\ \citenamefont {Garrahan}}]{Keys2015}%
  \BibitemOpen
  \bibfield  {author} {\bibinfo {author} {\bibfnamefont {A.~S.}\ \bibnamefont
  {Keys}}, \bibinfo {author} {\bibfnamefont {D.}~\bibnamefont {Chandler}}, \
  and\ \bibinfo {author} {\bibfnamefont {J.~P.}\ \bibnamefont {Garrahan}},\
  }\href@noop {} {\bibfield  {journal} {\bibinfo  {journal} {Phys. Rev. E}\
  }\textbf {\bibinfo {volume} {92}},\ \bibinfo {pages} {022304} (\bibinfo
  {year} {2015})}\BibitemShut {NoStop}%
\bibitem [{\citenamefont {Sasa}(2010)}]{Sasa2010}%
  \BibitemOpen
  \bibfield  {author} {\bibinfo {author} {\bibfnamefont {S.-i.}\ \bibnamefont
  {Sasa}},\ }\href@noop {} {\bibfield  {journal} {\bibinfo  {journal} {Journal
  of Physics A: Mathematical and Theoretical}\ }\textbf {\bibinfo {volume}
  {43}},\ \bibinfo {pages} {465002} (\bibinfo {year} {2010})}\BibitemShut
  {NoStop}%
\bibitem [{\citenamefont {Garrahan}(2014)}]{Garrahan2014}%
  \BibitemOpen
  \bibfield  {author} {\bibinfo {author} {\bibfnamefont {J.~P.}\ \bibnamefont
  {Garrahan}},\ }\href {\doibase 10.1103/PhysRevE.89.030301} {\bibfield
  {journal} {\bibinfo  {journal} {Phys. Rev. E}\ }\textbf {\bibinfo {volume}
  {89}},\ \bibinfo {pages} {030301} (\bibinfo {year} {2014})}\BibitemShut
  {NoStop}%
\bibitem [{\citenamefont {Jack}\ and\ \citenamefont
  {Garrahan}(2016)}]{Jack2016b}%
  \BibitemOpen
  \bibfield  {author} {\bibinfo {author} {\bibfnamefont {R.~L.}\ \bibnamefont
  {Jack}}\ and\ \bibinfo {author} {\bibfnamefont {J.~P.}\ \bibnamefont
  {Garrahan}},\ }\href {\doibase 10.1103/PhysRevLett.116.055702} {\bibfield
  {journal} {\bibinfo  {journal} {Phys. Rev. Lett.}\ }\textbf {\bibinfo
  {volume} {116}},\ \bibinfo {pages} {055702} (\bibinfo {year}
  {2016})}\BibitemShut {NoStop}%
\end{thebibliography}
%

\end{document}